\documentclass[times,twocolumn,final]{elsarticle}

\usepackage{cnf}
\usepackage{amsmath}
\usepackage{framed,multirow}
\usepackage{subcaption}

\usepackage{amssymb}
\usepackage{latexsym}

\usepackage{url}
\usepackage{xcolor}
\definecolor{newcolor}{rgb}{.8,.349,.1}

\usepackage{hyperref}

\usepackage[switch,pagewise]{lineno} 

\journal{Combustion \& Flame}

\begin{document}

\verso{Ashray Mohit et al.}

\begin{frontmatter}

\title{Comparison of the flamelet progress variable and thickened
flame models for LES of turbulent combustion}

\author[2]{Ashray \snm{Mohit}\corref{cor1}}
\cortext[cor1]{Corresponding author: Department of Mechanical Engineering, University of Michigan, Ann Arbor, 48109, USA}
\emailauthor{ashraym@umich.edu}{Ashray Mohit}
    
\author[2,3]{Margaret \snm{Wooldridge}}

\author[2,3]{Jesse \snm{Capecelatro}}

\address[2]{Department of Mechanical Engineering, University of Michigan, Ann Arbor, 48109, USA}
\address[3]{Department of Aerospace Engineering, University of Michigan, Ann Arbor, 48109, USA}

\begin{abstract}
Large-eddy simulation (LES) of turbulent combustion requires closure models for unresolved flame structures and chemical processes. Two widely used but fundamentally different approaches are the flamelet/progress-variable approach (FPVA) and the dynamically thickened-flame model (TFM). Here, both are applied to methane--air combustion within a common numerical framework and evaluated for the Sandia D flame and a jet-in-crossflow representative of a non-assist waste-gas flare across a range of crosswind-to-jet velocity ratios. For the predominantly diffusion-controlled Sandia-D flame, FPVA and TFM exhibit similar accuracy, while FPVA is substantially less computationally expensive and converges faster under mesh refinement. More detailed chemical kinetics improve agreement with experimental measurements, with little additional cost for FPVA but a substantial increase for TFM. For the jet-in-crossflow, both approaches reproduce the experimentally observed reduction in combustion efficiency with increasing crosswind, but their predictions progressively diverge as the crosswind strength increases. This divergence coincides with the emergence of a locally premixed reaction branch captured by TFM but not represented by FPVA, whereas predictions in diffusion-dominated regions remain broadly similar. In TFM, the locally premixed branch is associated with elevated scalar dissipation and greater residual methane downstream, indicating that highly strained premixed combustion contributes to differences in predicted flare performance. These results show that FPVA and TFM can provide similar predictions in diffusion-dominated flames but differ substantially when crossflow induces locally premixed, highly strained combustion near the nozzle.
\end{abstract}


\end{frontmatter}


\section*{Novelty and significance statement}

Turbulent flames in many applications span multiple combustion regimes, ranging from diffusion-dominated to locally premixed conditions, yet combustion closures are often assessed within a single regime. Consequently, the relative performance of different LES combustion closures in flows containing multiple local combustion regimes remains poorly understood. This study addresses this gap by comparing two widely used but fundamentally different LES combustion models, the flamelet/progress-variable approach (FPVA) and dynamically thickened-flame model (TFM), within a common numerical framework. The two approaches yield similar predictions in diffusion-dominated regions but diverge as increasing crosswind generates a locally premixed reaction branch. FPVA offers substantially lower computational cost and faster convergence under mesh refinement when diffusion-controlled combustion dominates, whereas TFM captures the emerging locally premixed branch that is not represented by the non-premixed FPVA manifold.

\section{Introduction}
\label{sec:introduction}
Large-eddy simulation (LES) has emerged as an important tool for predicting turbulent reacting flows in configurations for which direct numerical simulation is prohibitively expensive~\cite{pitsch2006large}. LES resolves the energetic turbulent motions responsible for large-scale transport and mixing, while the smaller turbulent and chemical scales remain unresolved. Combustion introduces an additional closure challenge because chemical source terms depend nonlinearly on temperature and composition, and physical reaction zones can be substantially thinner than the LES filter width. Accurately representing the resulting turbulence--chemistry interactions is therefore central to predictive LES of turbulent combustion.

Flamelet-based approaches address this problem by representing the local thermochemical state on a lower-dimensional manifold~\cite{peters1984laminar}. The flamelet/progress-variable approach (FPVA), introduced by Pierce and Moin~\cite{PierceMoin2004}, uses mixture fraction $Z$ and a reaction progress variable $C$ as manifold coordinates, enabling detailed chemistry without transporting every chemical species. In this approach, the expensive chemical-kinetics calculations are performed during offline table generation rather than in the simulation itself. Subsequent studies demonstrated that FPVA can capture local extinction, reignition, and strongly unsteady non-premixed flames \cite{IhmePitsch2008b}.

These computational savings rely on the assumption that the thermochemical state is uniquely determined by $Z$ and $C$. This assumption may break down in locally premixed regions, whose states may not be represented by the manifold. Previous studies have examined the limitations of representing premixed, partially premixed, and diffusion-flame structures within a common manifold. Fiorina et al.~\cite{Fiorina2005} showed that flamelet tabulation performs well as long as the local state remains within the regime represented by the library, but its accuracy degrades when diffusion across mixture-fraction surfaces becomes important, leading to errors in reaction-zone structure and species concentrations. Bray et al.~\cite{Bray2005} showed that partially premixed combustion introduces additional scalar-dissipation terms in the progress-variable formulation that are required to recover the diffusion-flame limit. Knudsen and Pitsch~\cite{KnudsenPitsch2009,KnudsenPitsch2012} demonstrated that errors can arise even when the local combustion regime is correctly identified, if the flamelet manifold does not adequately represent the full range of thermochemical states encountered by the flow. Thus, accurate identification of the combustion regime alone does not guarantee accurate predictions when the underlying manifold is incomplete.

An alternative strategy is to retain finite-rate chemistry and modify the physical reaction zone so that it can be represented on the LES mesh. Artificially thickened-flame models (TFM) increase diffusive transport while reducing chemical reaction rates such that the flame thickness increases without changing the corresponding laminar propagation speed~\cite{Colin2000,Charlette2002a}. Because the chemistry is evaluated from the transported thermochemical state, such approaches are not inherently restricted to a precomputed chemical manifold, but they require additional species transport and finite-rate chemistry integration.

The distinction between these modeling strategies becomes important when premixed and non-premixed reaction zones coexist within the same flow, as observed in simulations of lifted turbulent jet flames \cite{YooSankaranChen2009}. Multi-regime flamelet models have been developed to address this behavior by selecting or blending between premixed and non-premixed manifolds \cite{KnudsenPitsch2009,KnudsenPitsch2012}. However, these approaches increase modeling complexity because they require multiple manifolds, additional control variables, and procedures for selecting or blending between them, while errors can still arise when the local thermochemical state falls outside the tabulated space~\cite{nguyen2018impacts}. This is particularly relevant to reacting jets in crossflow, where strong mixing, strain, and recirculation can drive rapid transitions between combustion regimes. Kuenne et al.~\cite{Kuenne2011} combined flamelet-generated manifolds with artificial flame thickening, enabling premixed flames to be resolved on LES meshes while retaining an efficient treatment of detailed chemical kinetics through tabulation. Such hybrid approaches motivate independent assessment of FPVA and TFM to isolate their respective strengths and limitations and inform future coupled formulations.

Waste-gas flares provide a practical setting for assessing the influence of combustion modeling on LES predictions, particularly combustion efficiency. Crosswinds promote mixing, recirculation, and localized premixing, affecting flame stabilization and methane leakage. Smith et al.~\cite{Smith2013} were among the first to demonstrate that LES can reproduce measured flare combustion behavior under varying crosswind conditions. Using Bayesian inference, they showed that uncertainty bounds widened with increasing crosswind velocity. Spinti et al.~\cite{Spinti2022} developed a LES framework for a multi-flare setup and performed simulations for varying levels of fidelity, showing that far-field mesh refinement increased predicted combustion efficiency from 92.0\% to 94.5\%, while interactions between neighboring flare plumes had negligible effects on combustion efficiency under the conditions examined. Baranger and Poinsot~\cite{baranger2024stabilization} identified crosswind velocity and burner-lip temperature as key controls on flame anchoring, with cooler lips promoting lift-off and methane escape. Gaipl et al.~\cite{gaipl2025combustion} identified rich and lean premixed branches alongside diffusion combustion and showed that strain-sensitive chemistry altered predicted combustion efficiency by as much as 7\%. These studies highlight key mechanisms governing flare performance, but the influence of combustion-model choice, particularly FPVA versus TFM, remains largely unexplored.


The objective of this work is to quantitatively compare FPVA and TFM for LES of turbulent combustion, focusing on the Sandia-D flame and methane jets in crossflow. Both models are evaluated within a common LES framework (Sec.~\ref{sec:2}). Verification and validation are performed using the Sandia-D flame (Sec.~\ref{sec:sandiaD}), assessing grid sensitivity, computational cost, and mean and RMS temperature quantities against experiments. The models are then applied to a methane jet-in-crossflow (Sec.~\ref{sec:JICF}), comparing combustion efficiency, methane slip, and computational cost. Differences are interpreted through combustion regimes, scalar dissipation, and flame strain (Secs.~\ref{sec:prembranch} and~\ref{sec:state_strain_comparison}) to identify the mechanisms governing methane leakage and assess model applicability for flare simulations.

\section{Model formulation}
\label{sec:2}
The combustion models are evaluated within a large-eddy simulation (LES) framework for variable-density, low-Mach-number reacting flows, where the filtered continuity and momentum equations are
\begin{equation}
\frac{\partial \overline{\rho}}{\partial t}+\nabla\cdot
\left(
\overline{\rho}\widetilde{\mathbf{u}}
\right)
=0,
\label{eq:les_continuity}
\end{equation}
\begin{equation}
\begin{split}
\frac{\partial \overline{\rho}\widetilde{\mathbf{u}}}{\partial t}
+&
\nabla\cdot
\left(
\overline{\rho}
\widetilde{\mathbf{u}}
\otimes
\widetilde{\mathbf{u}}
\right)=-\nabla\overline{p}\\
+&
\nabla\cdot
\left[
(\mu+\mu_t)
\left(
\nabla\widetilde{\mathbf{u}}
+
\nabla\widetilde{\mathbf{u}}^{T}
-\frac{2}{3}
(\nabla\cdot\widetilde{\mathbf{u}})\mathbf{I}
\right)
\right],
\label{eq:les_momentum}
\end{split}
\end{equation}
where $\overline{\rho}$ is the density, $\widetilde{\mathbf{u}}$ is the velocity vector, $\overline{p}$ is the pressure, $\mathbf{I}$ is the identity tensor, $\mu$ is the molecular viscosity and $\mu_t$ is the turbulent viscosity closed using the dynamic Smagorinsky model \cite{lilly1992proposed}. Here, the overbar denotes a filtered quantity and the tilde denotes a Favre-filtered quantity.

The corresponding filtered species equation is
\begin{equation}
\frac{\partial \overline{\rho}\widetilde{Y}_k}{\partial t}
+
\nabla\cdot
\left(
\overline{\rho}
\widetilde{\mathbf{u}}
\widetilde{Y}_k
\right)
=
\nabla\cdot\left[\overline{\rho}\left(D_k+D_t\right)\nabla\widetilde{Y}_k\right]+\overline{\dot{\omega}_k},
\label{eq:les_species}
\end{equation}
where $\widetilde{Y}_k$ denotes the mass fraction of species $k$, $D_k$ is its molecular diffusivity, and $D_t=\mu_t/(\overline{\rho} {\rm Sc}_t)$ is the turbulent diffusivity with ${\rm Sc}_t=0.9$ the turbulent Schmidt number. The principal remaining closure is the filtered chemical source term, $\overline{\dot{\omega}_k}$, whose treatment distinguishes the two combustion models considered here. FPVA obtains the thermochemical state (temperature, density, and species mass fractions) and corresponding source terms from a precomputed flamelet manifold, whereas TFM directly evaluates finite-rate chemistry. In TFM, an additional equation is solved for enthalpy and density is determined from the equation of state. The diffusive fluxes and source terms are then modified to thicken the flame and enable the reaction zone to be resolved on the LES mesh. Further details on both approaches are provided below.

\subsection{Flamelet/progress-variable approach}
\label{sec:fpva_method}

\begin{figure}[!t]
\centering
\includegraphics[scale=0.18]{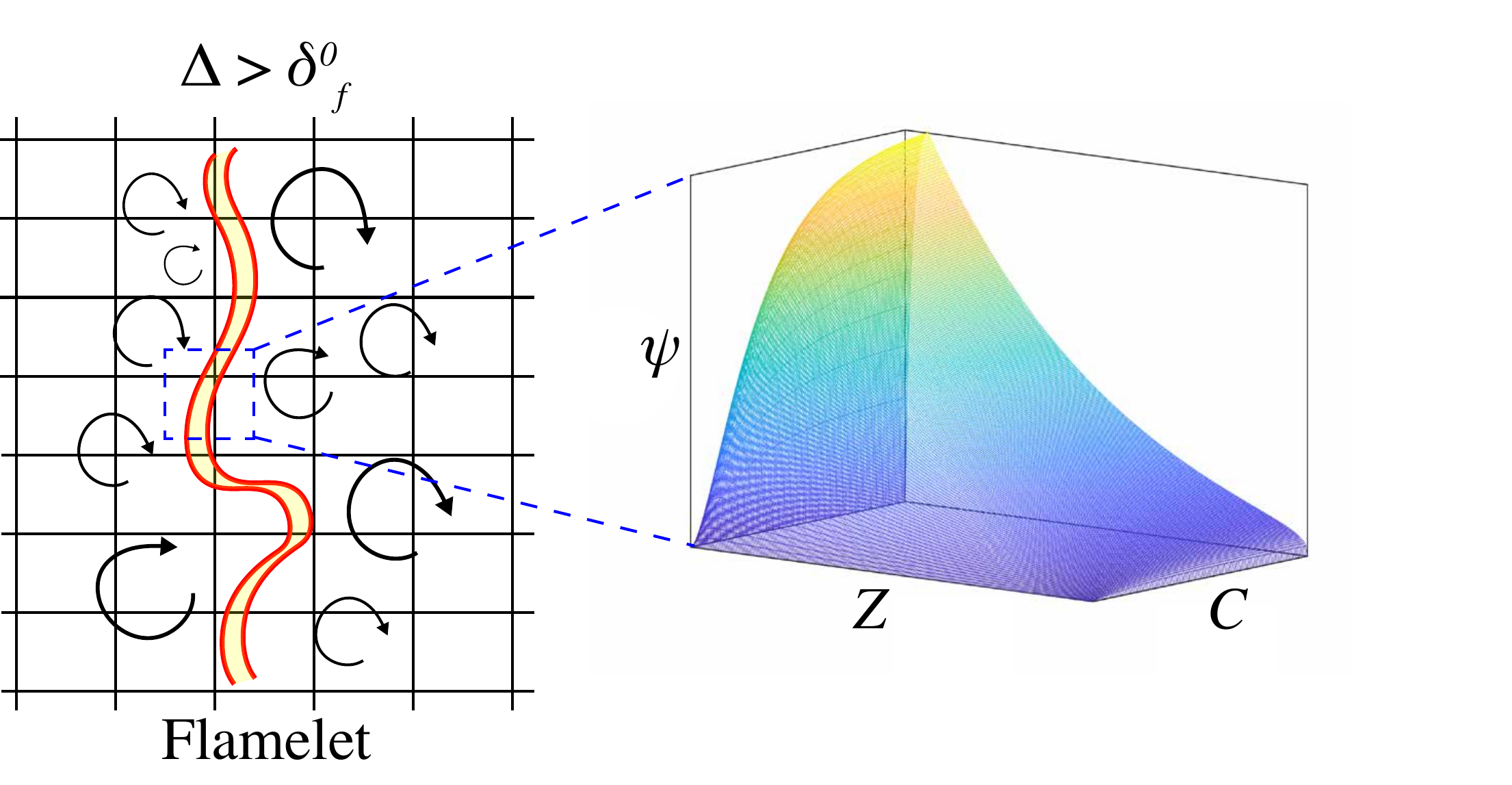}
\caption{Schematic of FPVA. The thermochemical state, $\psi$, is represented by a precomputed manifold parameterized by the mixture fraction $Z$ and progress variable $C$. The flame thickness $\delta_f^0$ is assumed to be asymptotically thin relative to the LES grid spacing, such that $\delta_f^0 \ll \Delta$.}
\label{fig:fpva_schematic}
\end{figure}

The flamelet formulation assumes that the internal chemical structure
of a turbulent flame can be represented by an ensemble of laminar
flamelets. For non-premixed combustion, the thermochemical state is parameterized by the mixture fraction, $Z$, which represents
the local mixing between the fuel and oxidizer streams. To represent
departure from chemical equilibrium, an additional progress variable,
$C$, is introduced. It is generally defined as a weighted sum of major
product species, $C=\sum_{k}a_kY_k$,
where the coefficients $a_k$ are selected such that $C$ increases
monotonically from the unburned to the burned state. Under the FPVA assumption, any thermochemical quantity $\psi$ is represented by a manifold, $\psi = \mathcal{M}_{\psi}(Z,C)$, where $\psi$ may denote temperature, density, species mass fraction, heat-release rate, transport properties, or a chemical source term.
A schematic representation of FPVA is shown in
Fig.~\ref{fig:fpva_schematic}.

The manifold is generated prior to the simulation by solving a series of one-dimensional flamelet equations over the relevant ranges of mixture fraction, scalar dissipation rate, and reaction progress~\cite{PierceMoin2004,peters_turbulent_combustion_2000}. The resulting thermochemical states are stored in a lookup table.

To access this manifold during LES, transport equations are solved
only for the reduced set of variables required to identify the local
thermochemical state. Rather than solving an equation for every
chemical species, FPVA advances the manifold coordinates and the
statistical quantities required by the presumed-PDF closure. The primary
coordinate is the mixture fraction, which is governed by
\begin{equation}
\frac{\partial\overline{\rho}\widetilde{Z}}{\partial t}
+
\nabla\cdot
\left(
\overline{\rho}\widetilde{\mathbf{u}}\widetilde{Z}
\right)
=
\nabla\cdot
\left[
\overline{\rho}
\left(
D_Z+D_t
\right)
\nabla\widetilde{Z}
\right],
\label{eq:fpva_Z}
\end{equation}
where $D_Z$ is the mixture-fraction diffusivity.
Because $\widetilde{Z}$ is constructed from conserved elemental mass
fractions, it contains no chemical source term. Chemical evolution along
the manifold is instead described through the progress variable, whose
transport equation is
\begin{equation}
\frac{\partial\overline{\rho}\widetilde{C}}{\partial t}
+
\nabla\cdot
\left(
\overline{\rho}\widetilde{\mathbf{u}}\widetilde{C}
\right)
=
\nabla\cdot
\left[
\overline{\rho}
\left(
D_C+D_t
\right)
\nabla\widetilde{C}
\right]
+
\overline{\dot{\omega}_C},
\label{eq:fpva_C}
\end{equation}
where $D_C$ is the progress variable diffusivity. Here, $\overline{\dot{\omega}_C}$ is obtained directly from the
precomputed flamelet library according to
\begin{equation}
\overline{\dot{\omega}_C}
=
\int_0^1
\int_0^1
\dot{\omega}^{\mathrm{tab}}_C(Z,C)
\widetilde{P}(Z,C)
\,\mathrm{d}C\,\mathrm{d}Z,
\label{eq:fpva_source_lookup}
\end{equation}
where $\dot{\omega}^{\mathrm{tab}}$ is the tabulated source term obtained from the precomputed manifold. A beta distribution is used to describe sub-filter fluctuations in
mixture fraction, $\widetilde{Z^{\prime\prime 2}}$, while fluctuations in progress variable are
computed via a Dirac delta distribution ($\delta$):
\begin{equation}
\widetilde{P}(Z,C)
=
\beta
\left(
Z;
\widetilde{Z},
\widetilde{Z^{\prime\prime 2}}
\right)
\delta
\left(
C-\widetilde{C}
\right).
\label{eq:fpva_pdf}
\end{equation}
Thus, FPVA does not require transport equations and source-term evaluations for every chemical species. Thermochemical quantities are reconstructed from the table using $\widetilde{Z}$, $\widetilde{C}$, and $\widetilde{Z^{\prime\prime 2}}$, given by
\begin{equation}
\widetilde{\psi}
=
\int_0^1
\int_0^1
\mathcal{M}_{\psi}(Z,C)
\widetilde{P}(Z,C)
\,\mathrm{d}C\,\mathrm{d}Z.
\label{eq:fpva_filtered_lookup}
\end{equation}


In the present implementation, the library is constructed from non-premixed flamelets. Consequently, premixed or partially premixed reaction branches that are not represented by the manifold and cannot be recovered solely by increasing LES resolution.

\subsection{Dynamically thickened-flame model}
\label{sec:dtf_method}

The dynamically thickened-flame model resolves the reaction zone on
the LES mesh by artificially increasing its thickness while explicitly
transporting the retained chemical species and evaluating finite-rate
chemical source terms during the simulation. The laminar flame speed and flame thickness scale as
\begin{equation}
S_L^0\sim\left(\frac{\alpha}{\tau_c}\right)^{1/2},
\qquad
\delta_L^0\sim\left(\alpha\tau_c\right)^{1/2},
\label{eq:laminar_scaling}
\end{equation}
where $\alpha$ is the thermal diffusivity and $\tau_c$ is the chemical
time scale. Introducing a thickening factor, $F$, such that
\begin{equation}
D_k^\star=FD_k,
\qquad
\dot{\omega}_k^\star=\frac{1}{F}\dot{\omega}_k,
\label{eq:classical_thickening}
\end{equation}
in Eq.~\eqref{eq:les_species} increases the flame thickness to $F\delta_L^0$ while preserving the
laminar flame speed~\cite{Colin2000}.

The classical thickened-flame scaling is derived from premixed-flame
behavior and is therefore applied selectively in the present formulation.
In particular, artificial thickening is restricted to locally premixed
reaction zones. Applying the same transformation within a
diffusion-dominated flame would modify molecular transport and the
mixture-fraction gradients that control the local scalar dissipation rate,
potentially altering the physical structure of the diffusion flame
\cite{Colin2000,Cuenot2022}.

To distinguish between the local combustion regimes, the Takeno flame
index (TI) is employed \cite{Yamashita1996,Zirwes2021}. TI is based on
the relative alignment of the fuel and oxidizer mass-fraction gradients via
\begin{equation}
\mathrm{TI}
=
\frac{
\nabla Y_F \cdot \nabla Y_{O}
}{
\left|\nabla Y_F\right|
\left|\nabla Y_{O}\right|
},
\label{eq:takeno_index}
\end{equation}
providing a local indicator of premixed and diffusion-dominated
combustion. A regime indicator, $\mathcal{I}_{\mathrm{pm}}=(1+\rm{TI})/2$ is then
constructed such that
\begin{equation}
\mathcal{I}_{\mathrm{pm}}
=
\begin{cases}
1,  & \text{locally premixed combustion},\\
0, & \text{diffusion-dominated combustion}.
\end{cases}
\label{eq:premixed_indicator}
\end{equation}
The local thickening factor therefore depends on both the reaction-zone
sensor and the combustion regime,
\begin{equation}
\begin{aligned}
F
&=
1+\left(F_s-1\right)\Omega\,\mathcal{I}_{\mathrm{pm}},
\\
F_s
&=
\min
\left[
F_{\max},
\max\left(
1,\frac{N_{\Delta}\Delta}{\delta_L^0}
\right)
\right].
\end{aligned}
\label{eq:dynamic_thickening}
\end{equation}
where $\Delta$ is the local grid spacing, $N_{\Delta}$ is the desired
number of cells across a premixed flame front, and $F_{\max}$ limits
excessive thickening. The reaction sensor $\Omega\in[0,1]$ restricts
thickening to the reacting layer, while $\mathcal{I}_{\mathrm{pm}}$
restricts its application to regions identified as locally premixed.

\begin{figure}[!t]
\centering
\includegraphics[scale=0.18]{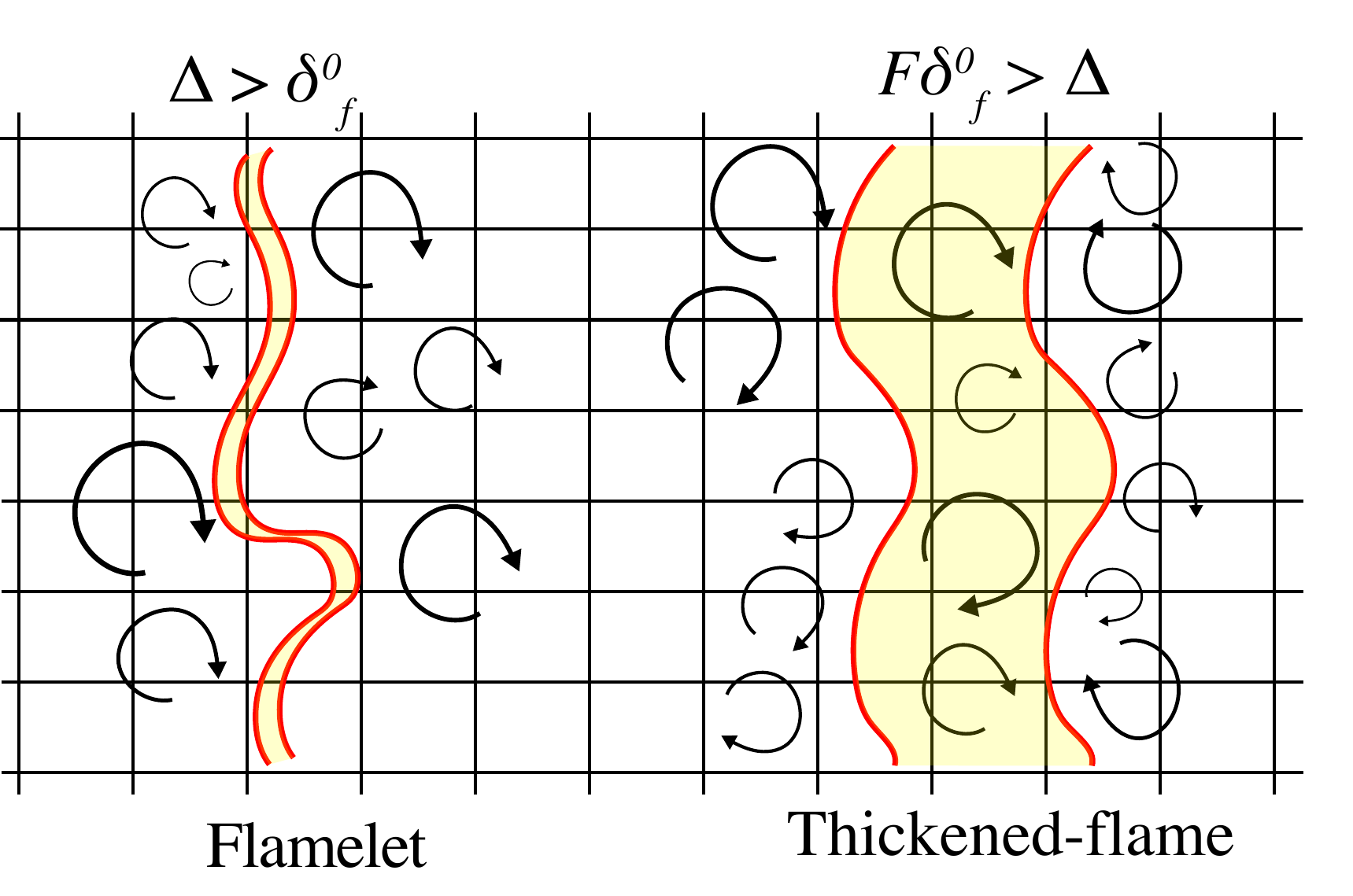}
\caption{Schematic of the thickened-flame approach. TFM artificially increases the flame thickness $\delta_f^0\rightarrow F\delta_f^0$, allowing the reaction zone to be resolved on the LES mesh.}
\label{fig:tfm_schematic}
\end{figure}

Consequently, $F>1$ only where a premixed reaction zone is insufficiently
resolved. In diffusion-dominated regions,
$\mathcal{I}_{\mathrm{pm}}=0$ and therefore $F=1$. The mesh size should be chosen
such that these diffusion-flame regions are resolved directly, allowing
the transported species and finite-rate chemistry to determine the local
flame structure without artificial modification of molecular transport.
The premixed flame thickening treatment is illustrated in
Fig.~\ref{fig:tfm_schematic}.

Although thickening preserves the planar laminar flame speed, it
reduces the ability of turbulent eddies to wrinkle the flame. An
efficiency function, $E$, is therefore introduced to compensate for
the unresolved flame surface. The molecular diffusion and chemical
source terms become
\begin{equation}
D_k^\star=EFD_k,
\qquad
\dot{\omega}_k^\star=\frac{E}{F}\dot{\omega}_k.
\label{eq:dtf_transform}
\end{equation}
Consequently, the effective flame speed and thickness scale as
\begin{equation}
S_L^\star \sim E S_L^0,
\qquad
\delta_L^\star \sim F\delta_L^0.
\label{eq:dtf_scaling}
\end{equation}
Thus, $F$ controls the numerical flame thickness, whereas $E$ accounts
for the increase in burning rate associated with unresolved turbulent
flame wrinkling.

The resulting equation for the Favre-filtered mass fraction of species $k$ is
\begin{align}
\frac{\partial\overline{\rho}\widetilde{Y}_k}{\partial t}
+
\nabla\cdot
\left(
\overline{\rho}\widetilde{\mathbf{u}}\widetilde{Y}_k
\right)
={}&
\nabla\cdot
\left\{
\overline{\rho}
\left(
EFD_k+D_t
\right)
\nabla\widetilde{Y}_k
\right\}
\nonumber\\
&+
\frac{E}{F}
\dot{\omega}_k
\left(
\widetilde{T},
\overline{\rho},
\widetilde{Y}_k
\right).
\label{eq:dtf_species}
\end{align}
The same
thickening transformation is applied consistently to the enthalpy equation.

The efficiency function is evaluated using the Charlette power-law
flame-wrinkling model \cite{Charlette2002a}. The
wrinkling factor associated with a laminar flame of thickness $\delta_L^0$ is
\begin{equation}
\Xi_{\Delta_e}(\delta_L^0)
=
\left\{
1+
\min
\left[
\frac{\Delta_e}{\delta_L^0},
\Gamma_{\Delta_e}
\frac{u'_{\Delta_e}}{S_L^0}
\right]
\right\}^{\beta_C},
\label{eq:charlette_wrinkling}
\end{equation}
where $\Delta_e$ is the effective scale, $u'_{\Delta_e}$ is the
unresolved turbulent velocity fluctuation, and
$\Gamma_{\Delta_e}$ represents the efficiency with which unresolved
eddies wrinkle the flame. The two terms in the minimum operator limit
wrinkling according to the available range of unresolved length scales
and the strength of the unresolved turbulence, respectively.

The efficiency function is defined as the ratio between the wrinkling
of the physical and artificially thickened flames:
\begin{equation}
E
=
\frac{
\Xi_{\Delta_e}\left(\delta_L^0\right)
}{
\Xi_{\Delta_e}\left(F\delta_L^0\right)
},
\qquad
\Delta_e=F\delta_L^0.
\label{eq:charlette_efficiency}
\end{equation}
This formulation corrects only the flame surface lost through
artificial thickening and ensures that $E=1$ when $F=1$. The present
implementation employs the fixed Charlette power-law exponent
$\beta_C=0.5$. Unlike FPVA, the thickened-flame model is not restricted
to a precomputed thermochemical manifold, allowing premixed, partially
premixed, and diffusion-dominated combustion modes to arise directly
from the transported species and finite-rate chemistry. When the Takeno indicator suppresses
thickening, $F=1$ and, from Eq.~\eqref{eq:charlette_efficiency},
$E=1$. The modified transport and reaction terms therefore reduce to
their physical values, $D_k^\star=D_k$ and $\dot{\omega}_k^\star=\dot{\omega}_k$.
Thus, TFM acts as a thickened-flame model only within locally premixed
regions, while diffusion-dominated combustion is treated using resolved
finite-rate chemistry. For brevity in the coming sections, bars and tildes are omitted for all quantities.

\subsection{Numerical setup}
\label{sec:numerical_setup}

Simulations are performed using custom solvers developed within the
OpenFOAM framework \cite{Weller1998,Hassanaly2018}. The solvers employ a low-Mach-number, variable-density formulation on a
collocated, fully unstructured mesh composed of tetrahedral cells.
Spatial discretization is second-order accurate, and a fractional-step procedure is used for pressure projection. The
convective terms in the momentum equation are evaluated using a
skew-symmetric formulation to minimize artificial kinetic-energy
dissipation, which is particularly important for LES where excessive
numerical dissipation can interfere with the resolved turbulent energy
cascade~\cite{Morinishi2010,Hassanaly2018}. 
For the TFM formulation, viscosity is determined using a temperature-dependent power law, while thermodynamic properties are evaluated using NASA polynomials~\cite{mcbride2002nasa}. For all simulations in this study, the same numerical methods are used for FPVA and TFM. This ensures that differences in predicted flame structure and global performance are primarily attributable to the combustion closure rather than the numerical treatment.

\subsection{Chemical kinetics}
Unless otherwise specified, the methane--air chemistry is represented using the two-step $\mathrm{BFER}{*}$ mechanism~\cite{Franzelli2013}. The same chemistry is used for both combustion closures. In FPVA, it is used to construct the flamelet library, whereas in TFM, the finite-rate reaction rates are evaluated directly during the LES. This ensures that differences between the two approaches primarily reflect the combustion closure rather than the chemical mechanism.

The original BFER mechanism was developed for premixed methane--air combustion and calibrated primarily using laminar premixed flame properties over a broad range of operating conditions~\cite{Franzelli2012}. It was subsequently applied to lean, partially premixed turbulent combustion representative of gas-turbine combustors~\cite{Franzelli2012,Franzelli2013}. The modified $\mathrm{BFER}{*}$ mechanism improves the sensitivity to flame stretching by adjusting species transport and reaction-rate parameters~\cite{Cuenot2022,Franzelli2013}. This is important for the present configuration, where jet shear and crosswind interaction generate strongly strained reaction zones, particularly near the flame base.Flamelet libraries for FPVA are generated using FlameMaster~\cite{Flamemaster1998} for methane--air counterflow diffusion flames, spanning the full mixture-fraction range ($0 \leq Z \leq 1$), with scalar dissipation rates varied up to the extinction limit.

\section{Results}
In this section, the effects of FPVA and TFM are evaluated in simulations of the canonical Sandia-D flame and a reacting methane jet-in-crossflow representative of a non-assisted waste-gas flare.



\subsection{Sandia-D flame}
\label{sec:sandiaD}
The Sandia-D flame is a well-characterized turbulent methane--air flame
with extensive experimental data for model validation. Its predominantly
diffusion-controlled structure provides a useful baseline for comparing
FPVA and TFM under conditions where similar predictions are expected.


The burner consists of a central fuel jet of diameter $D=7.2$~mm surrounded
by an annular pilot and an air coflow. The central jet contains 25\% methane
and 75\% air by volume and enters at a bulk velocity of $U_j=49.6$~m/s,
corresponding to a Reynolds number ${\rm Re}\approx22{,}400$. The pilot stream enters at
$U_p=11.4$~m/s and consists of a lean burned mixture corresponding to an equivalence ratio,
$\phi=0.77$ at approximately $1880$~K, while the surrounding air coflow
enters at $U_c=0.9$~m/s \cite{jaravel2018prediction}. The computational domain
extends approximately $35D$ downstream and $11D$ across.

The unstructured mesh is strongly refined in the central jet, pilot shear layers, and reaction zone to resolve steep velocity and scalar gradients, with progressively coarser cells used away from the flame to reduce computational cost. Velocity perturbations are prescribed at the fuel and pilot inlets to represent inlet turbulence~\cite{PitschSteiner2000}. The outer boundaries are
treated as outlets with a clipped Neumann condition on velocity, while the jet and pilot surfaces employ no-slip conditions. Pressure and temperature are treated with a zero-gradient Neumann conditions.


\begin{figure}[ht]
\centering
    \includegraphics[scale=0.28]{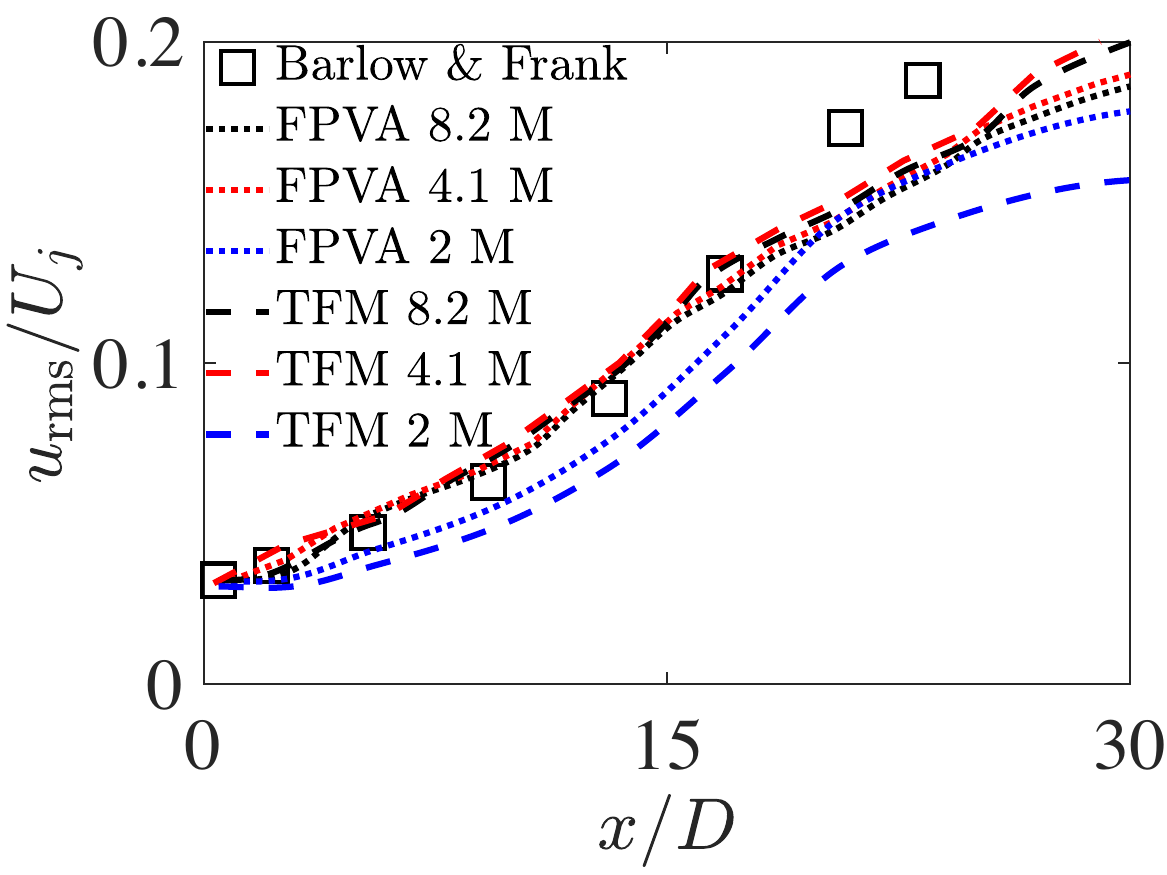}
    \caption{Mesh-sensitivity of the time-averaged normalized axial velocity
fluctuation, $u_{\mathrm{rms}}/U_j$, for the Sandia-D flame using FPVA and
TFM. Experimental
pmeasurements of Barlow and Frank~\cite{BarlowFrank1998} are included for comparison.}
\label{fig:sandia_mesh_convergence}
\end{figure}

Figure~\ref{fig:sandia_mesh_convergence} shows the sensitivity of the predicted velocity fluctuations to mesh resolution using approximately
2.0, 4.1, and 8.2 million elements. All quantities shown are time-averaged after the flow
reaches a statistically stationary state. Statistics are accumulated over
20 flow-through times, where the flow-through time is defined as $\tau_{\mathrm{FT}}=35D/U_j$. For both FPVA and TFM, refinement from 2.0 to 4.1 million elements produces a noticeable change in velocity fluctuatons, $u_{\mathrm{rms}}$, particularly farther downstream where the turbulent
jet has developed. In contrast, the 4.1 and 8.2 million element cases
remain comparatively close over most of the domain, indicating that the
solution becomes substantially less sensitive to further refinement and therefore the 4.1 million element mesh is used for all further comparisons in this section. It can also be seen that TFM exhibits larger deviation on the coarsest mesh compared with FPVA.

\begin{figure}[ht]

    \includegraphics[scale=0.27]{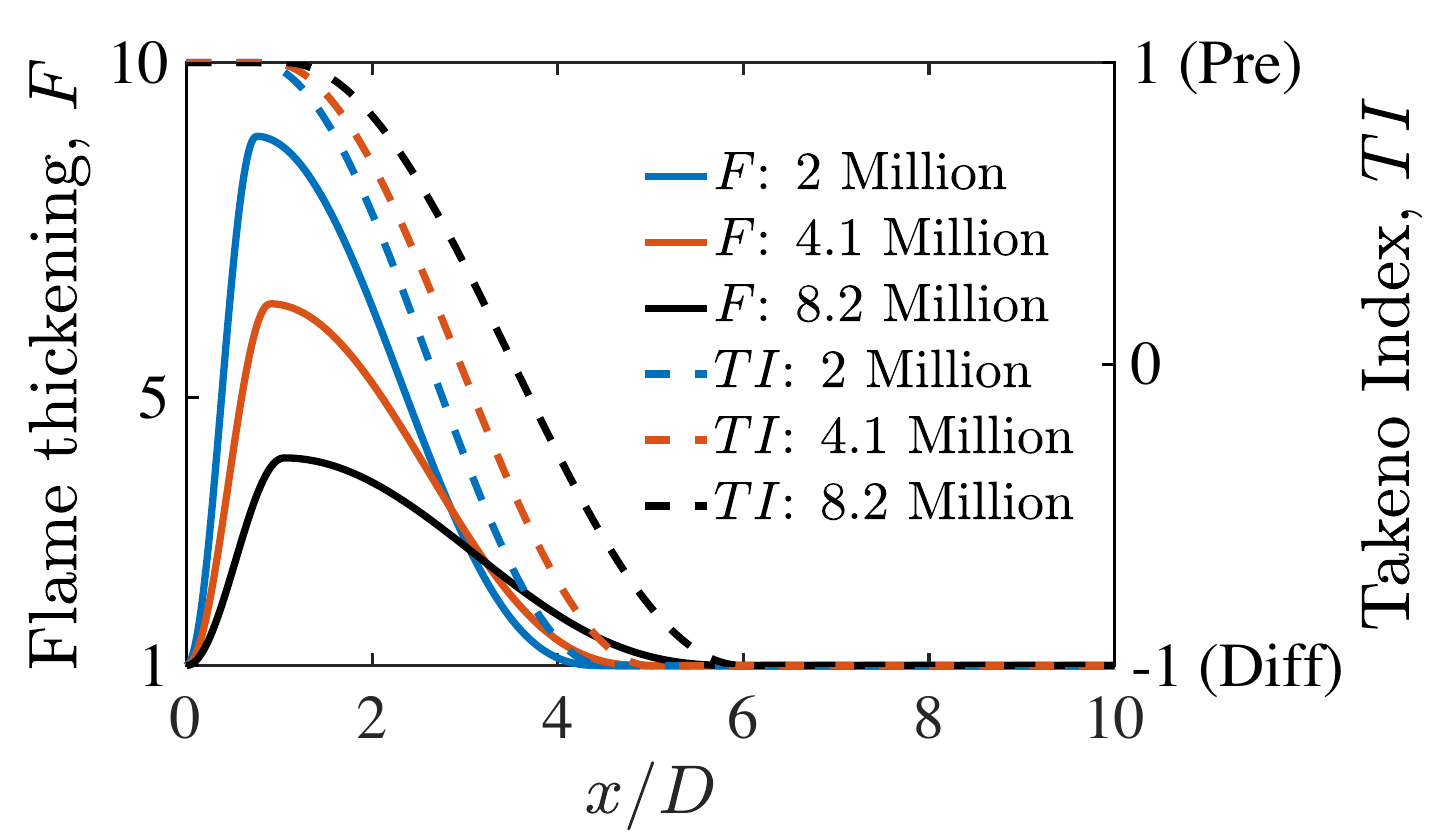}
\caption{Streamwise evolution of the flame-thickening factor $F$ and
Takeno Index for the TFM simulations at three mesh resolutions. Here, \textit{Diff} refers to the diffusion branch and \textit{Pre} refers to the diffusion branch.}
\label{fig:tfm_thickening}
\end{figure}

For the TFM cases, mesh refinement also directly reduces the amount of
artificial flame thickening required. As shown in
Fig.~\ref{fig:tfm_thickening}, the peak thickening factor decreases
systematically as the mesh is refined from approximately 2.0 to
8.2 million elements. This behavior is expected from the model
formulation, since the required thickening scales with the local grid
spacing. On the finest mesh, the physical flame is better resolved and
the model therefore approaches the unthickened limit, $F\rightarrow 1$,
over a larger portion of the domain.

The Takeno Index profiles further show that the thickened region is
confined to the locally premixed portion of the flame. As the solution
transitions downstream toward diffusion-dominated combustion, the
thickening factor returns to unity and the chemistry is resolved without
artificial flame thickening.

\begin{figure}[!t]
    \includegraphics[scale=0.28]{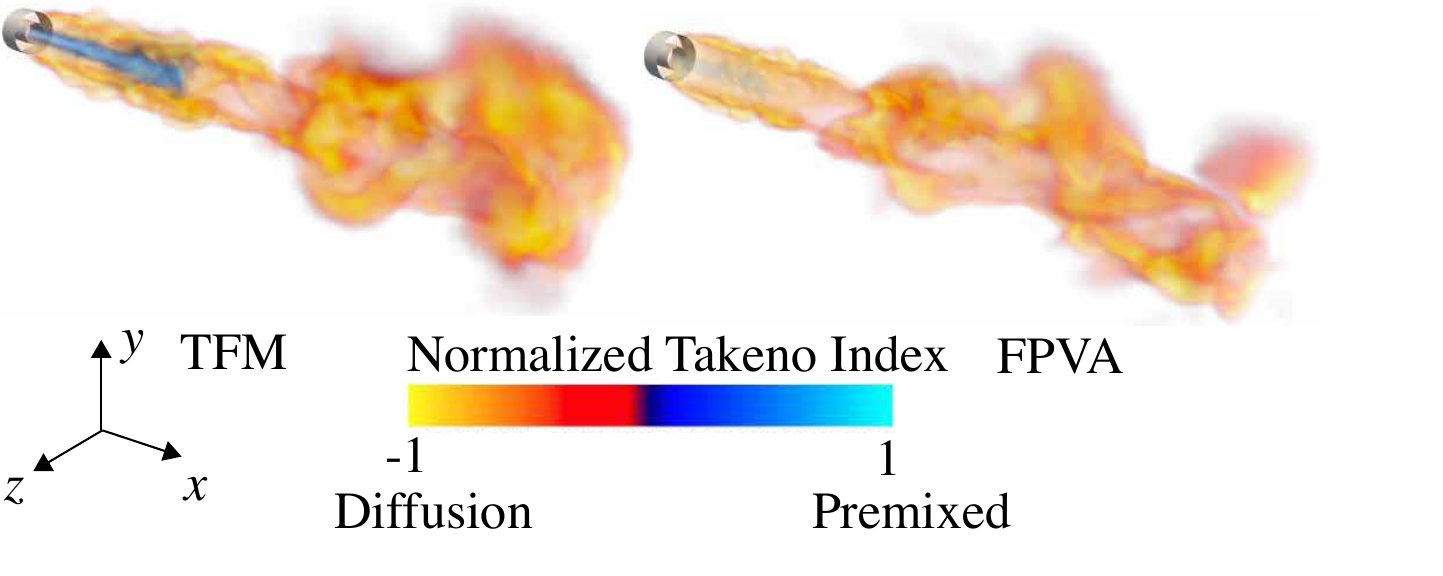}
    \caption{Instantaneous visualization of the combustion regime predicted by TFM
    (left) and FPVA (right), colored by the normalized Takeno Index.}
    \label{fig:regime_visualization}
\end{figure}

A qualitative comparison of the Sandia-D flame simulated using the two models is shown in Fig.~\ref{fig:regime_visualization}. TFM predicts a locally premixed region near the burner, whereas the FPVA solution remains diffusion-dominated. Farther downstream, both solutions become increasingly diffusion-controlled. The approaches therefore produce qualitatively similar downstream diffusion-flame structures but differ most strongly in the near field, where premixed combustion develops. To quantify these differences, one-dimensional profiles are presented at selected downstream locations to compare the local thermochemical structures predicted by the two models.

\begin{figure*}[!t]
    \centering


    \begin{subfigure}[t]{0.3\textwidth}

        \includegraphics[width=0.95\textwidth]{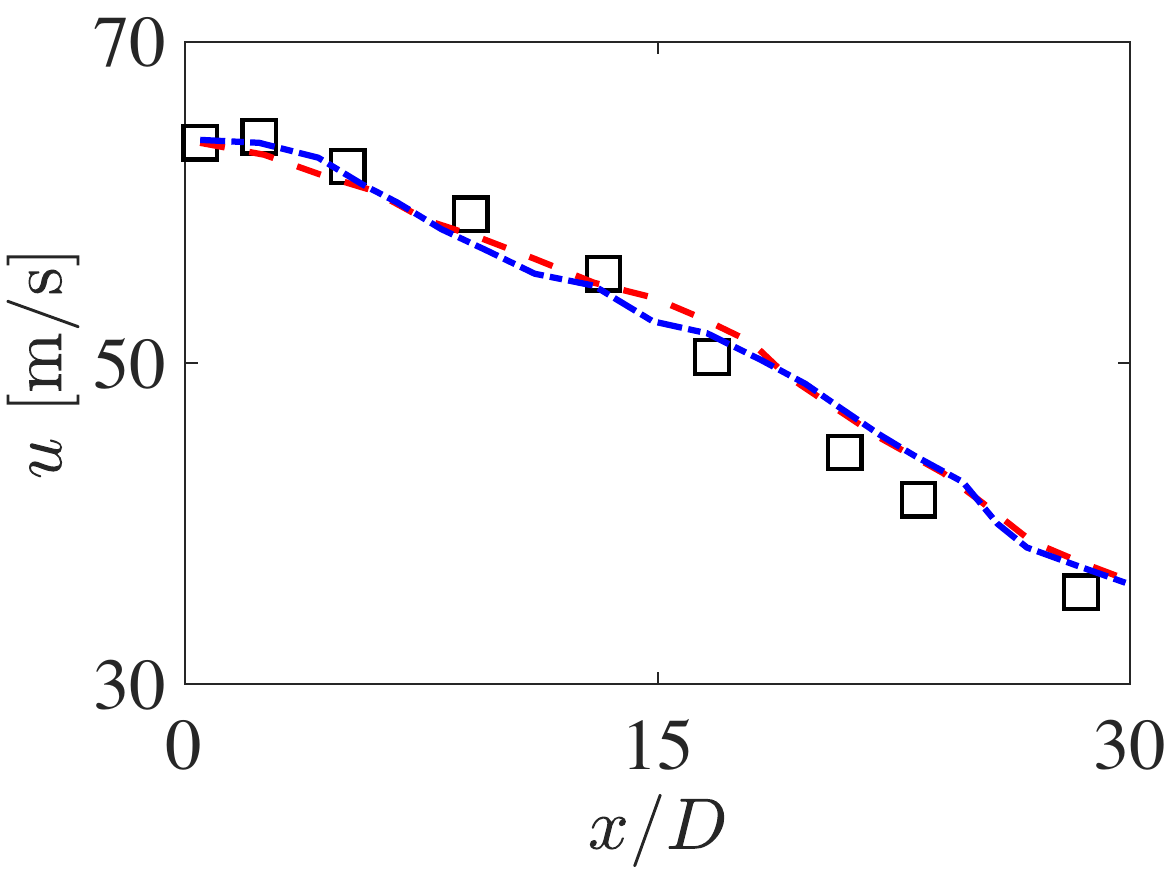}
        \caption{}
        \label{fig:sandia_urms}
    \end{subfigure}
    \hfill
    \begin{subfigure}[t]{0.3\textwidth}
        \includegraphics[width=0.95\textwidth]{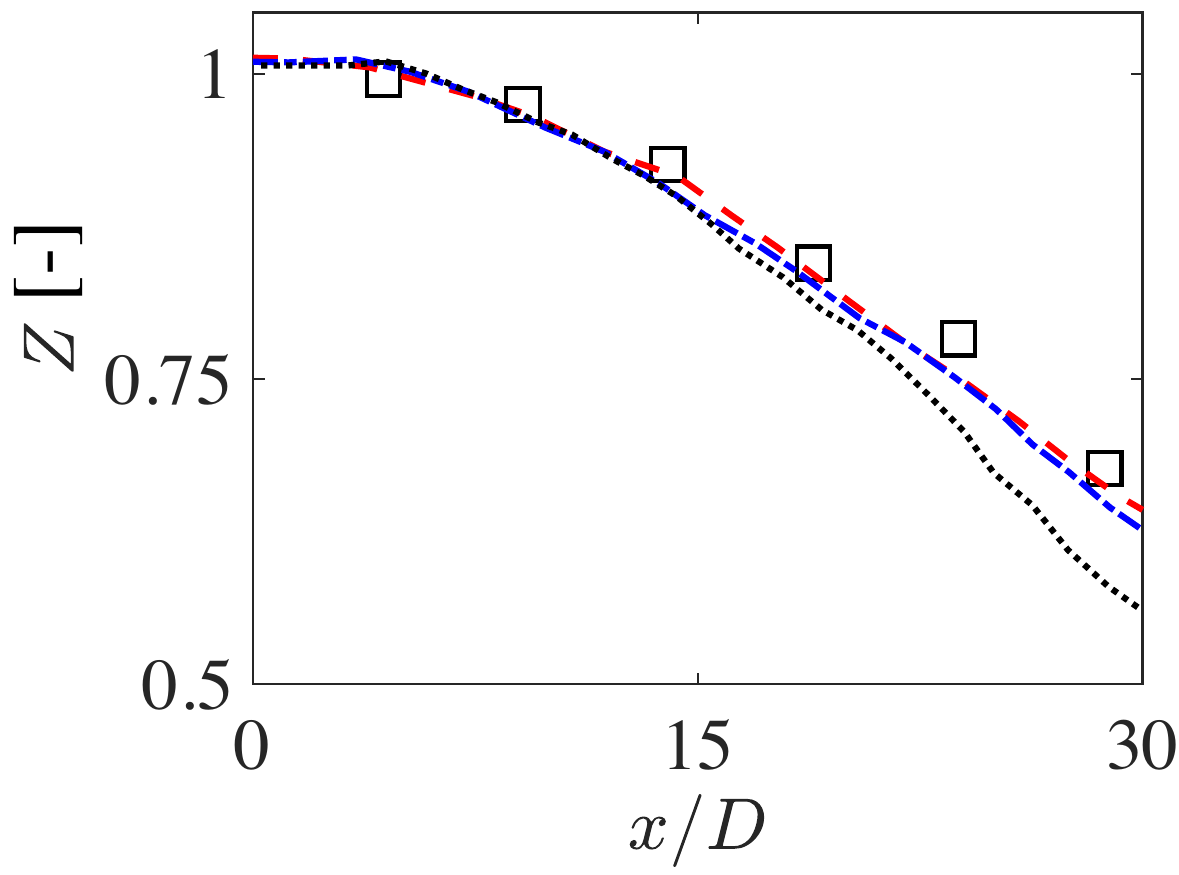}
        \caption{}
        \label{fig:sandia_zrms}
    \end{subfigure}
    \hfill
    \begin{subfigure}[t]{0.3\textwidth}
        \includegraphics[width=0.95\textwidth]{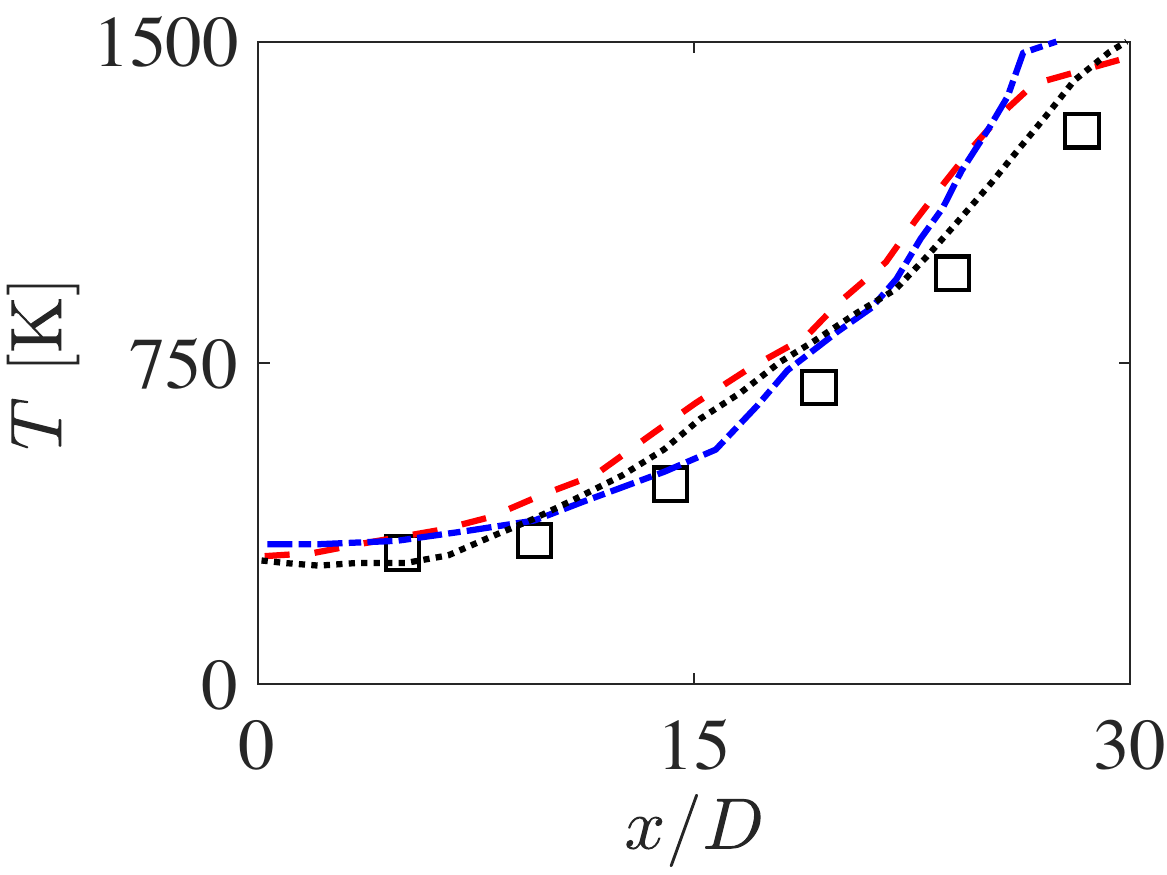}
        \caption{}
        \label{fig:sandia_trms}
    \end{subfigure}

    \vspace{0cm}


    \begin{subfigure}[t]{0.3\textwidth}
        \includegraphics[width=0.95\textwidth]{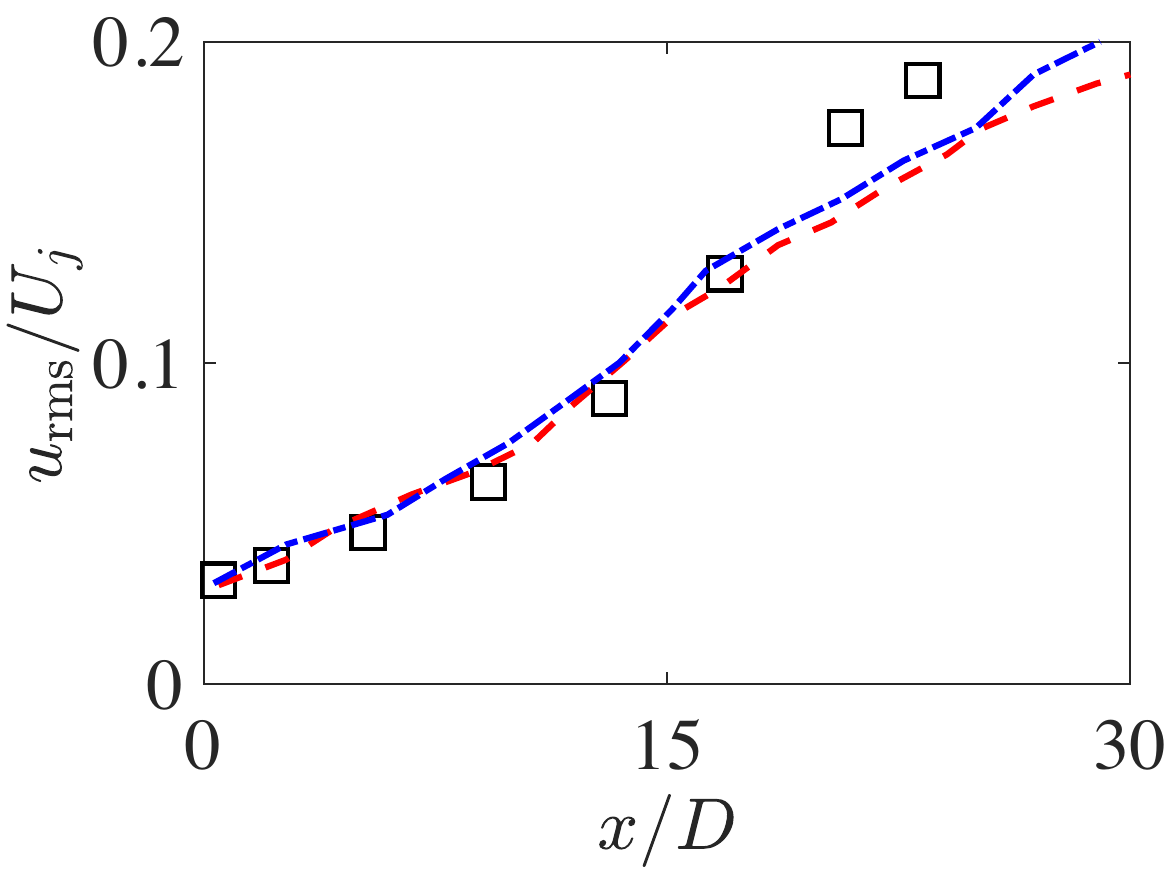}
        \caption{}
        \label{fig:sandia_umean}
    \end{subfigure}
    \hfill
    \begin{subfigure}[t]{0.3\textwidth}
        \includegraphics[width=0.95\textwidth]{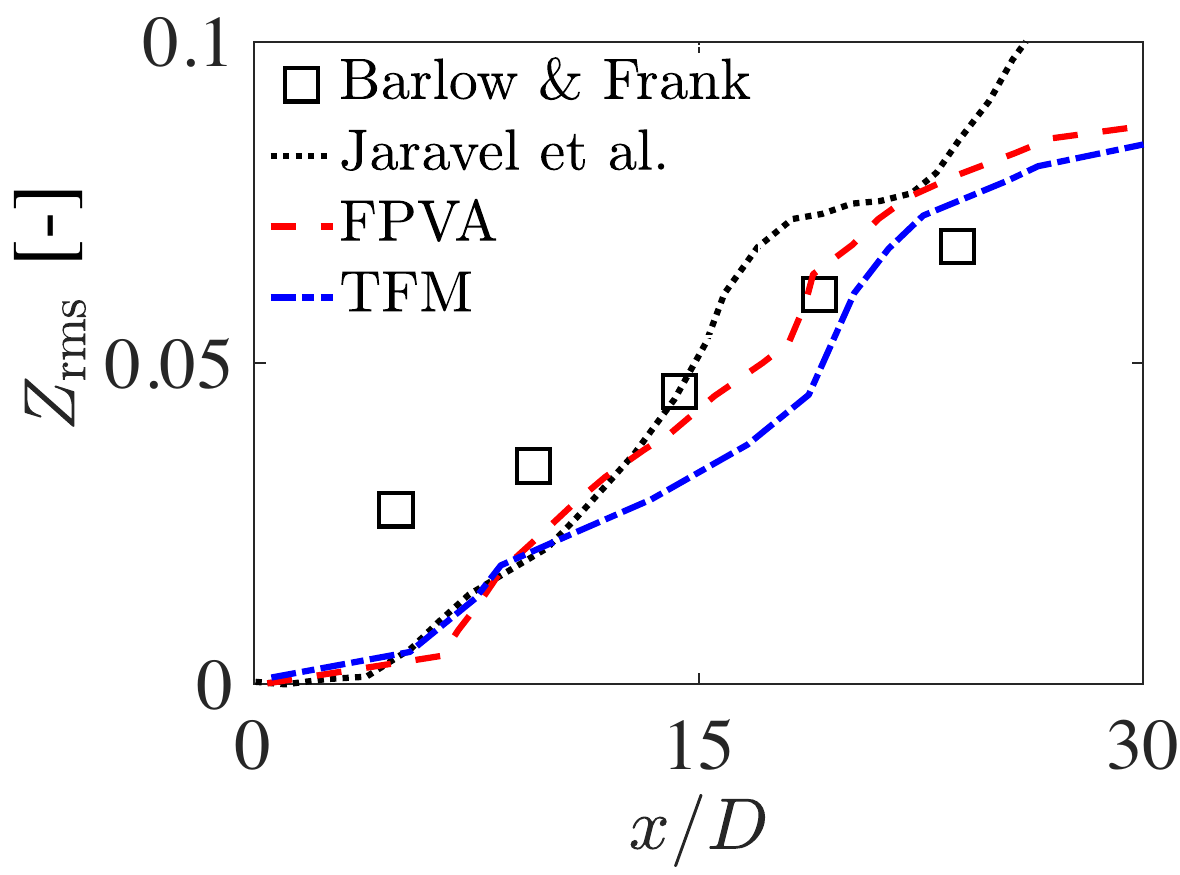}
        \caption{}
        \label{fig:sandia_zmean}
    \end{subfigure}
    \hfill
    \begin{subfigure}[t]{0.3\textwidth}
        \includegraphics[width=0.95\textwidth]{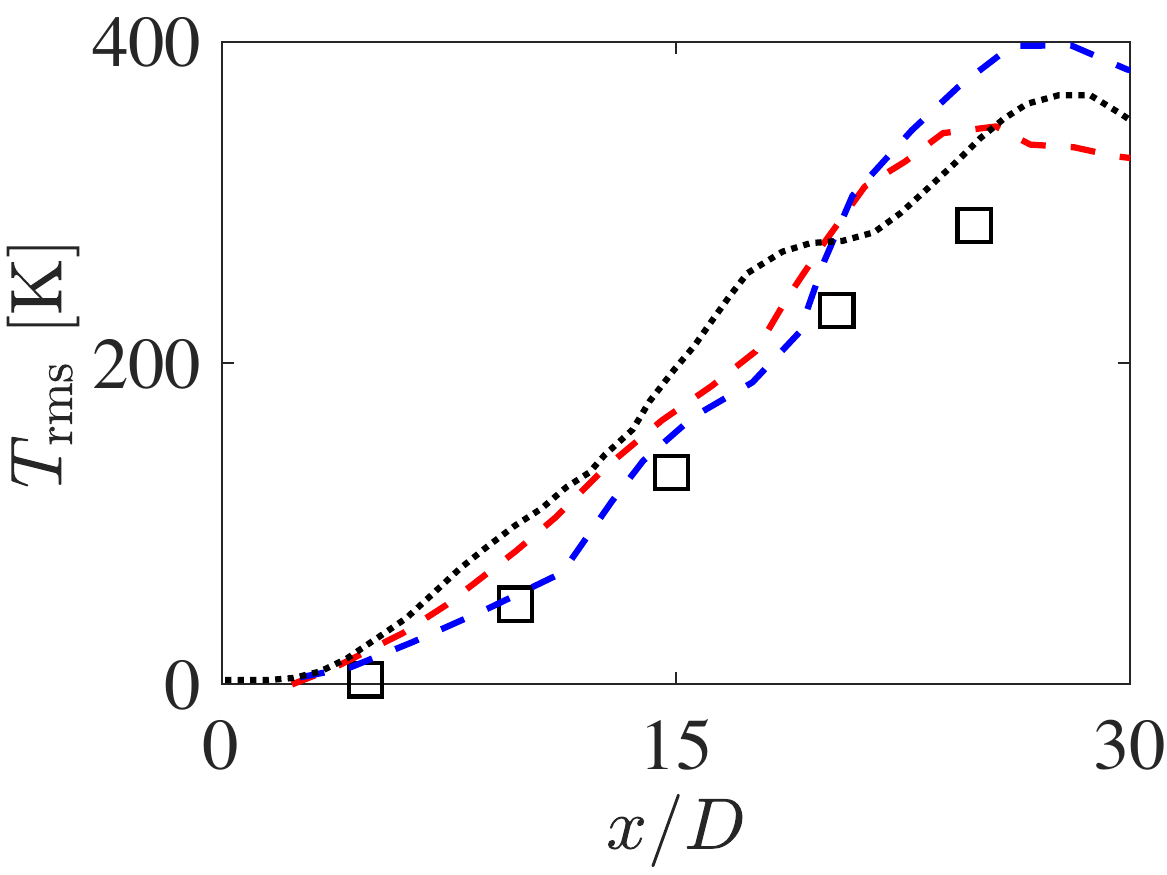}
        \caption{}
        \label{fig:sandia_tmean}

    \end{subfigure}

    \caption{Comparison of FPVA (red) and TFM (blue) for the Sandia-D flame. Mean quantities (top row), RMS fluctuations (bottom row). Experimental measurements by Barlow and Frank~\cite{BarlowFrank1998} (squares) and LES by Jaravel et al.~\cite{jaravel2018prediction} (black dotted line) shown for reference.}
    \label{fig:sandia_1d_profiles}

\end{figure*}

Figure~\ref{fig:sandia_1d_profiles} compares the centerline mean and root-mean-square (RMS) quantities predicted by FPVA and TFM with the experiments of Barlow and Frank\cite{BarlowFrank1998} and the LES results of Jaravel et al.~\cite{jaravel2018prediction}. Jaravel et al. directly integrated a 22-species chemical mechanism on a mesh containing 375 million tetrahedral elements and reported a computational cost of approximately $5\times10^5$ CPU-hours per flow-through time, providing an additional high-fidelity numerical reference. The mean axial velocity, mixture fraction, and temperature differ only slightly between FPVA and TFM, and both models reproduce the overall downstream evolution observed experimentally. The overall good agreement is consistent with the  diffusion-controlled character of the Sandia D flame, for which the two models represent similar local combustion physics.

Differences between the models are more apparent in the root-mean-square (RMS) quantities. Note the underprediction of $Z_{\rm rms}$ near the nozzle ($x=0$) likely results from the inlet boundary condition, because no mixture-fraction perturbations were imposed. Velocity fluctuations remain relatively similar between FPVA and TFM, whereas larger differences emerge in the mixture-fraction and, particularly, temperature fluctuations farther downstream. These quantities are more sensitive to differences in the instantaneous reaction-zone structure and turbulence--chemistry interactions, even when the corresponding mean fields remain similar. Nevertheless, both models capture the overall development of the turbulent flame, with predictions generally within the spread of the experimental measurements and previously reported numerical results. Despite yielding similar predictions for the Sandia D flame, TFM is substantially more expensive, requiring approximately 2.6 times as many CPU-hours per flow-through time as FPVA (see Table~\ref{tab:computational_cost}).

Additional simulations were performed using the one-step methane mechanism (1S$\_$CH4)~\cite{CERFACS1SCH4MP1} with both FPVA and TFM, and GRI-Mech 3.0~\cite{Smith1999GRI30} with FPVA. The one-step mechanism is a highly reduced model designed to reproduce key methane--air flame properties at low computational cost, but does not account for intermediate species or CO oxidation. In contrast, GRI-Mech 3.0 is a detailed methane mechanism with 53 species and 325 reactions, validated against a broad range of combustion measurements~\cite{Smith1999GRI30}. BFER* provides an intermediate level of chemical complexity. TFM was not run with GRI-Mech 3.0 due to the high computational cost. Table~\ref{tab:computational_cost} reports the normalized $L_2$ errors in $T_{\rm{rms}}$ relative to the experimental measurements (similar trends were observed in velocity and mixture fraction RMS). More detailed chemistry improves agreement with the experiments for both models. Although FPVA and TFM yield similar predictions with BFER*, FPVA is substantially less expensive because it retrieves the thermochemical state from a lookup table. Consequently, FPVA can accommodate increased chemical-kinetic detail with little additional LES runtime cost, whereas the corresponding cost increase for TFM is substantial.

\begin{table}[ht]
\caption{Computational cost for the Sandia-D simulations.}
\label{tab:computational_cost}
\centering
\fontsize{11pt}{11pt}\selectfont
\setlength{\tabcolsep}{4pt}
\renewcommand{\arraystretch}{1.15}

\resizebox{\columnwidth}{!}{%
\begin{tabular}{lcccc}
\hline
Model &
Chemistry &
Time per timestep [s] &
$L_2$ error ($T_{\rm{rms}}$) &
CPU h / $\tau_{\mathrm{FT}}$
\\
\hline

FPVA &
BFER* &
0.32 &
0.171 &
1000
\\

TFM &
BFER* &
0.91 &
0.183 &
2600
\\

FPVA &
1S\_CH4 &
0.32 &
0.278 &
1000
\\

TFM &
1S\_CH4 &
0.68 &
0.312 &
2000
\\

FPVA &
GRI-3.0 &
0.35 &
0.148 &
1200
\\

\hline
\end{tabular}%
}
\end{table}

\subsection{Jet-in-crossflow}
\label{sec:JICF}
\begin{figure}[t!]
    \centering
    \includegraphics[scale=0.25]{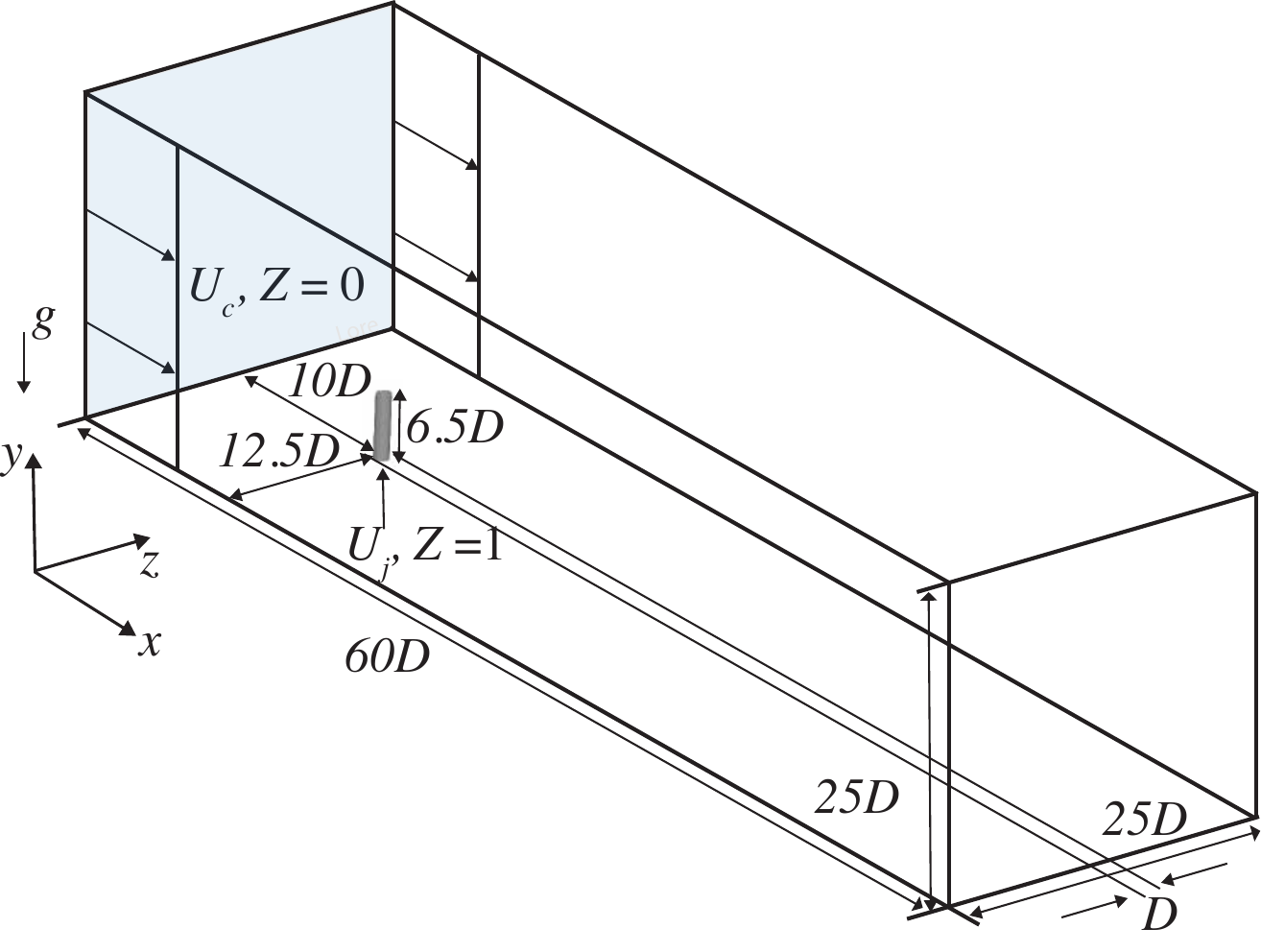}
    \caption{Computational configuration for the reacting jet-in-crossflow.}
    \label{fig:jicf_setup}
\end{figure}

The jet-in-crossflow (JICF) configuration shown in
Fig.~\ref{fig:jicf_setup} is used as a representative model for a
non-assist waste-gas flare. The computational domain extends $60D$ in the
streamwise direction and $25D$ in both the vertical and spanwise
directions. The nozzle is positioned $10D$ downstream of the crossflow
inlet and protrudes $6.5D$ into the domain, with the nozzle exit located at $x/D=0$. Unlike a conventional flush-mounted JICF, the elevated
nozzle better represents practical flare geometries while retaining the
characteristic downstream shear layers and counter-rotating vortex pair
of a jet-in-crossflow~\cite{Mohit2026,mohit2026effect}.

Pure methane enters the nozzle at $U_j=3$ m/s with $Z=1$, while air enters through the crossflow boundary at $3 \le U_c \le 12$ m/s with $Z=0$, corresponding to a velocity ratio $r=U_c/U_j$ in the range $1 \le r \le 4$. Dirichlet conditions are prescribed for velocity and transported scalars at both inlets, with zero-gradient pressure. Adiabatic, no-slip boundary conditions are imposed at the nozzle wall, with zero-gradient conditions for the scalars and pressure. All remaining boundaries are treated as outlets using a clipped-Neumann condition for velocity and zero-gradient conditions for the other variables.

\begin{figure*}[!t]
    \centering

    \begin{subfigure}[t]{0.7\textwidth}
        \centering
        \includegraphics[width=\textwidth]
        {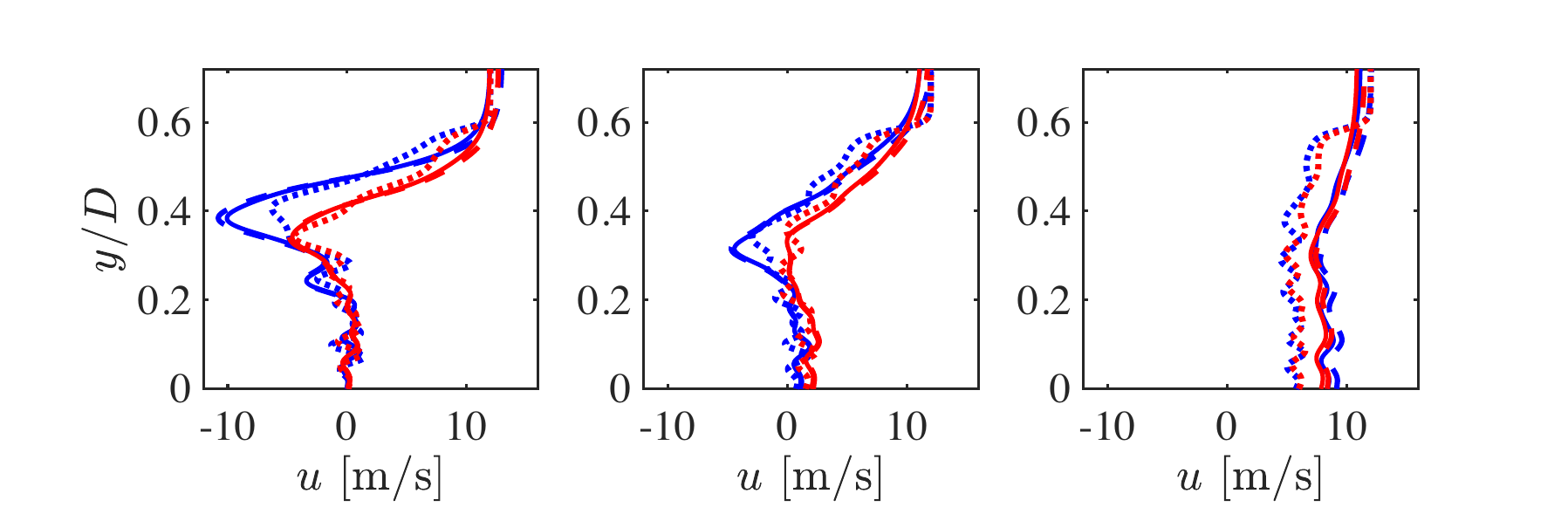}
        \caption{}
        \label{fig:jicf_velocity_profiles}
    \end{subfigure}

    \vspace{0cm}

    \begin{subfigure}[t]{0.7\textwidth}
        \centering
        \includegraphics[width=\textwidth]
        {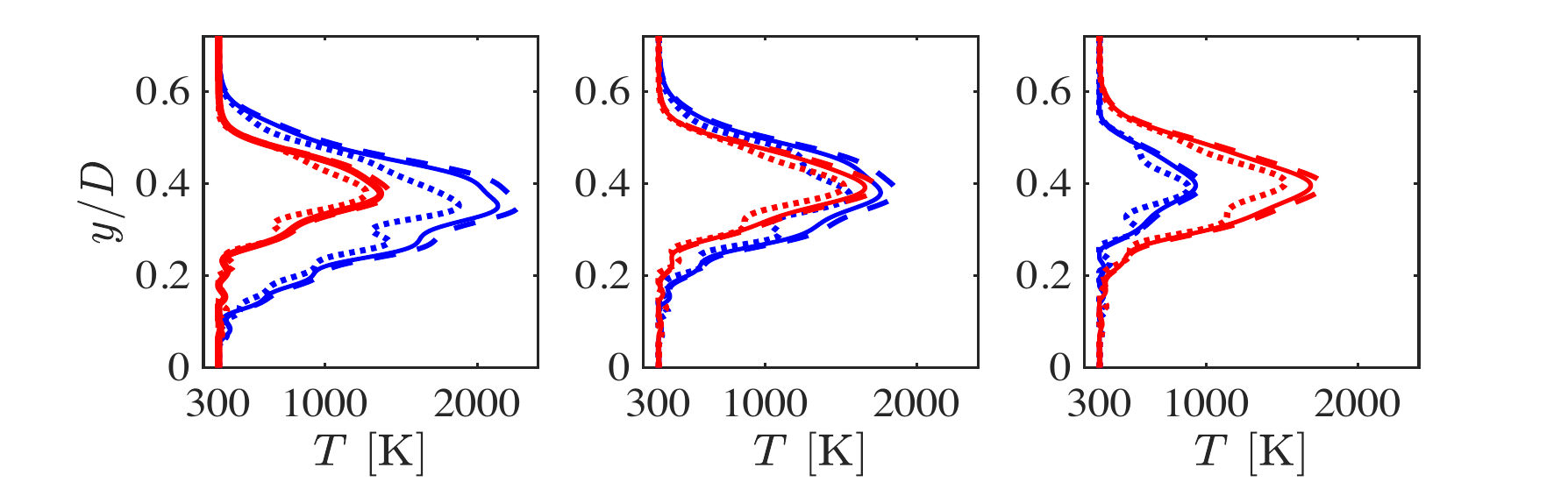}
        \caption{}
        \label{fig:jicf_temperature_profiles}
    \end{subfigure}

    \vspace{0cm}

    \begin{subfigure}[t]{0.7\textwidth}
        \centering
        \includegraphics[width=\textwidth]
        {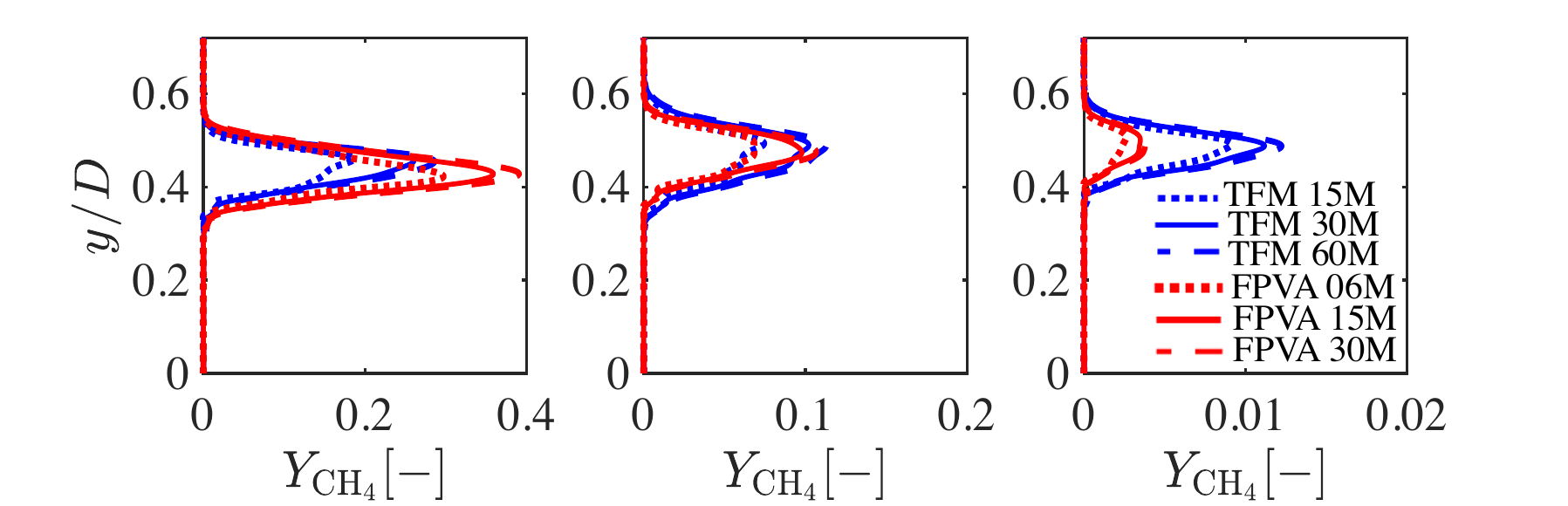}
        \caption{}
        \label{fig:jicf_ch4_profiles}
    \end{subfigure}

    \caption{Mesh-refinement comparison for the jet-in-crossflow at $r=4$. FPVA (red) and TFM (blue) predictions are shown in the $z=0$ plane at $x/D=1$ (left), $6$ (middle), and $12$ (right). From top to bottom, the rows show (a) time-averaged streamwise velocity, (b) temperature, and (c) methane mass fraction.}
    \label{fig:jicf_grid_model_comparison}

\end{figure*}

The influence of mesh resolution is first examined in Fig.~\ref{fig:jicf_grid_model_comparison} before the two combustion models are compared. Three refinement levels are considered for each model. The FPVA calculations use meshes containing approximately 6, 15, and 30 million elements, whereas the TFM calculations require substantially finer meshes containing approximately 15, 30, and 60 million elements. Unless otherwise stated, statistics are accumulated over 20 flow-through times to ensure statistical convergence. The flow-through time is defined using the axial domain length and crosswind velocity as $\tau_{\mathrm{FT}}=60D/U_c$.

For both models, mesh sensitivity is greatest in regions of strong velocity, temperature, and composition gradients. This is most evident at $x/D=1$, where the jet shear layer, recirculation zone, and reaction region are confined to a relatively small region. Successive refinement produces progressively smaller changes in the profiles, and the influence of grid resolution generally decreases farther downstream as the jet becomes more dispersed. The FPVA solutions differ only slightly between the 15- and 30-million-element meshes, indicating that the principal mean-flow and scalar structures are well represented at these resolutions.

In contrast, TFM exhibits a stronger dependence on mesh resolution,
particularly in the near-field and within the narrow reacting and mixing
layers. This higher resolution requirement follows from the manner in
which the chemistry is represented. Artificial thickening is applied only
within locally premixed regions, while diffusion-dominated portions of the
flame approach $F=1$ and are therefore resolved directly using finite-rate
chemistry. Consequently, adequate representation of the diffusion flame,
strong scalar gradients, and near-nozzle recirculation requires a finer
computational mesh. The 60-million-element TFM case reflects this greater spatial-resolution requirement.

Near the nozzle
($x/D=1$), TFM predicts a substantially stronger recirculation region,
as indicated by the larger negative streamwise velocities near the jet
shear layer. FPVA also captures the recirculation but underpredicts its magnitude relative to TFM. The difference remains apparent at $x/D=6$ but largely disappears by $x/D=12$, where the crossflow increasingly dominates the streamwise velocity field.

\begin{figure}[!t]
    \centering
    \includegraphics[scale=0.24]
    {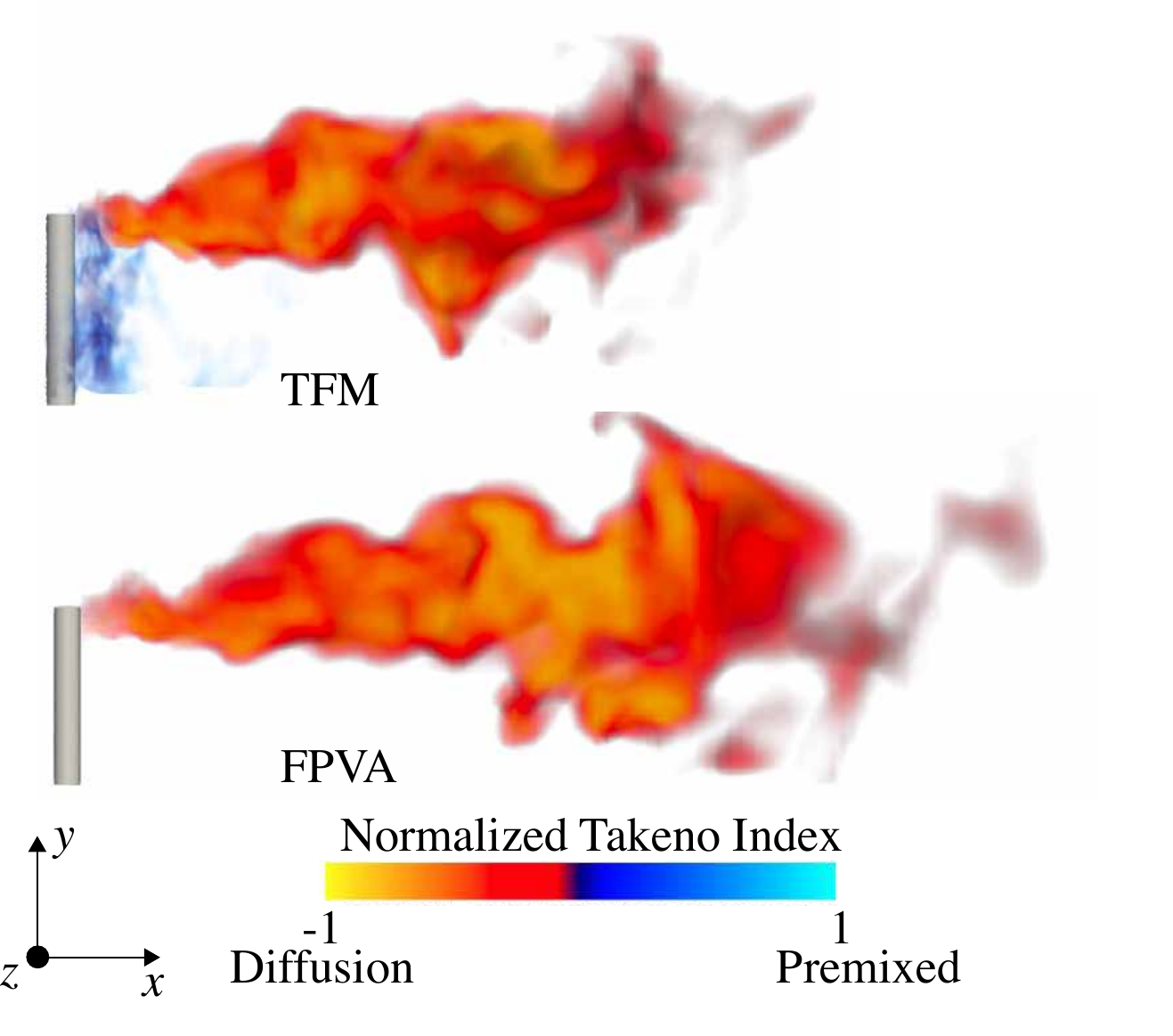}
    \caption{Instantaneous visualization of the combustion regime predicted by TFM
    (left) and FPVA (right), colored by the normalized Takeno Index for a JICF with $r=2$.}
    \label{fig:jicfviz}
\end{figure}

The temperature profiles also reveal differences in the reaction zone near the nozzle. At $x/D=1$, TFM predicts a considerably
stronger high-temperature region, indicating significant heat release
close to the nozzle. FPVA predicts a weaker near-field temperature rise,
consistent with the absence of the locally premixed reaction branch
observed in the visualization of the Takeno Index in Fig.~\ref{fig:jicfviz}. Farther downstream,
the temperature fields become more comparable at $x/D=6$, although
differences again develop in the far-field structure at $x/D=12$.

The methane profiles further demonstrate that differences generated in
the near-field persist downstream. At $x/D=1$, FPVA retains a somewhat
larger peak methane concentration, while TFM consumes more fuel within
the strongly reacting near-nozzle region. At $x/D=6$, the two solutions
are closer in magnitude but retain distinct profile shapes. By
$x/D=12$, however, TFM predicts a noticeably larger residual methane
concentration than FPVA. Thus, although TFM exhibits stronger
near-nozzle burning, this does not correspond to uniformly greater fuel
conversion throughout the plume; appreciable unburned methane remains
farther downstream. These differences in the spatial distribution of
methane ultimately motivate comparison of the integrated combustion
efficiency between the two models.

\begin{figure}[!t]
    \centering
    \includegraphics[scale=0.29]
    {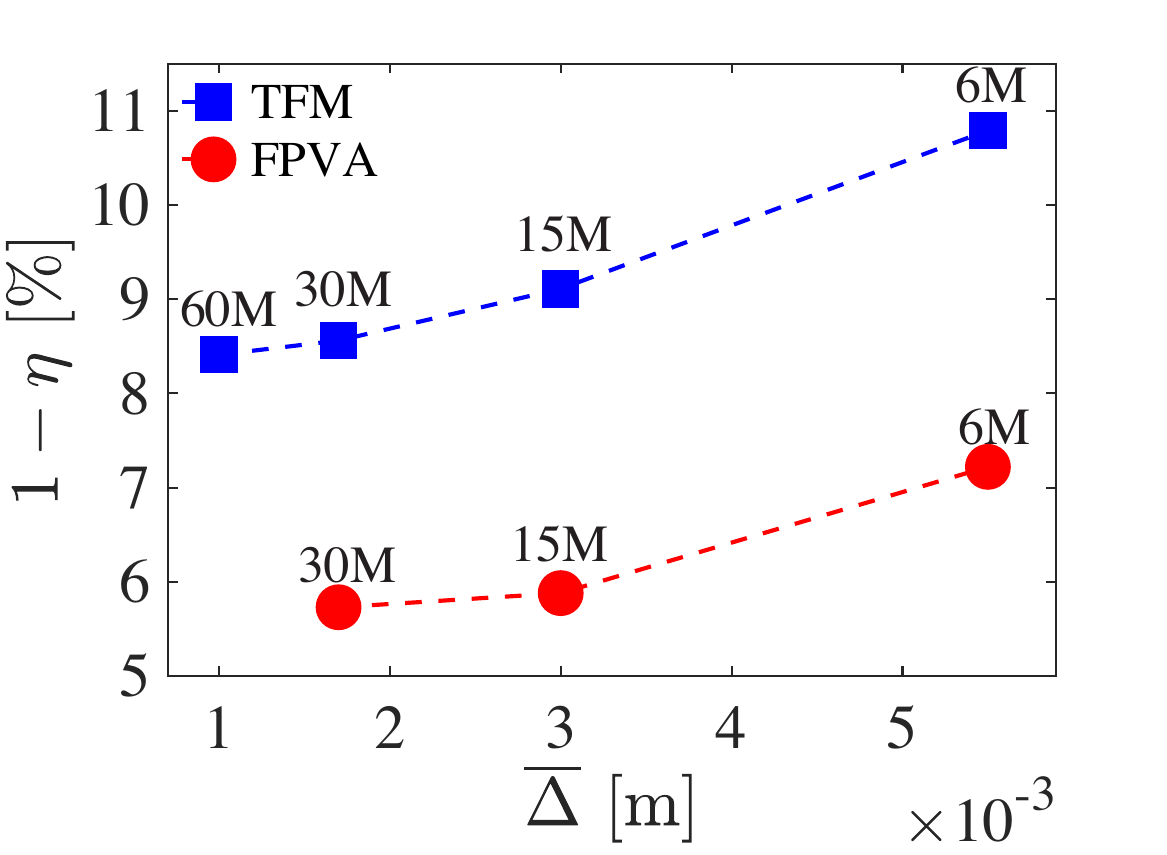}
    \caption{Effect of mesh refinement on the predicted combustion
    inefficiency, $1-\eta$, for FPVA (red) and TFM (blue) in the JICF with $r=4$. $\overline{\Delta}$ represents the average element size.}
    \label{fig:jicf_ce_grid}
\end{figure}

The effect of grid refinement on the integrated combustion efficiency is shown in Fig.~\ref{fig:jicf_ce_grid}. Combustion efficiency is evaluated from a global carbon balance and defined as the fraction of carbon entering through the fuel nozzle that exits the domain as carbon dioxide, given by
\begin{equation}
\eta
=
\frac{
\left\langle
\dot{m}_{\mathrm{CO_2}}
\right\rangle
}{
\dot{m}_{C,\mathrm{in}}
},
\label{eq:combustion_efficiency}
\end{equation}
where $\dot{m}_{C,\mathrm{in}}
=
\rho_j A_j U_j Y_{C}$ is the incoming carbon mass flow rate with $\rho_j$ the jet density, $A_j$ the nozzle area, $U_j$ the jet velocity, and $Y_{\rm C}$ the carbon mass fraction of the jet. $\dot{m}_{\mathrm{CO_2}}$ denotes the mass flow rate of $\mathrm{CO_2}$ through the downstream domain boundary. Angle brackets denote time averaging. The instantaneous outgoing carbon flux is evaluated as
\begin{equation}
\dot{m}_{\mathrm{CO_2}}(t)
=
\int_{A_E}
\rho_{\mathrm{CO_2}} Y_{\mathrm{CO_2}}
\left(
\mathbf{u}\cdot\hat{\mathbf{n}}
\right)
\,\mathrm{d}A_E,
\label{eq:co2_carbon_flux}
\end{equation}
where $A_E$ denotes the downstream sampling surface, 
$\mathbf{u}\cdot\hat{\mathbf{n}}$ represents the velocity component normal to the exit plane, and $Y_{\mathrm{CO_2}}$ is the mass fraction of carbon dioxide. Because the flare is unsteady, the outgoing carbon flux is averaged only
after statistical stationarity is reached, over 20 flow-through times
defined using the axial domain length and crosswind velocity. The downstream sampling location is chosen so that carbon conversion is
complete and $\eta$ is independent of position. Wooldridge et al.
\cite{Wooldridge2026CE} showed that sampling approximately $50D$
downstream reduces the associated bias to below $0.1\%$.

For both combustion models, decreasing
the average grid spacing $\overline{\Delta}$ reduces the predicted
combustion inefficiency, $1-\eta$, with progressively smaller changes
between the finest meshes. For FPVA, the inefficiency decreases from
approximately 7.2\% on the 6 million element mesh to approximately 5.7\%
on the 30 million element mesh. TFM exhibits a stronger resolution
requirement, decreasing from approximately 10.7\% on the 6 million
element mesh to approximately 8.4\% on the 60 million element mesh.
The comparatively small change between the finest two resolutions for
each model indicates that the global efficiency is approaching a
mesh-independent value. Importantly, even after the solutions
approach mesh convergence, TFM predicts a systematically larger
combustion inefficiency than FPVA. The remaining difference therefore
cannot be attributed solely to grid resolution and instead reflects
differences in the combustion physics represented by the two closures.

\begin{figure}[!t]
    \centering
    \includegraphics[scale=0.23]
    {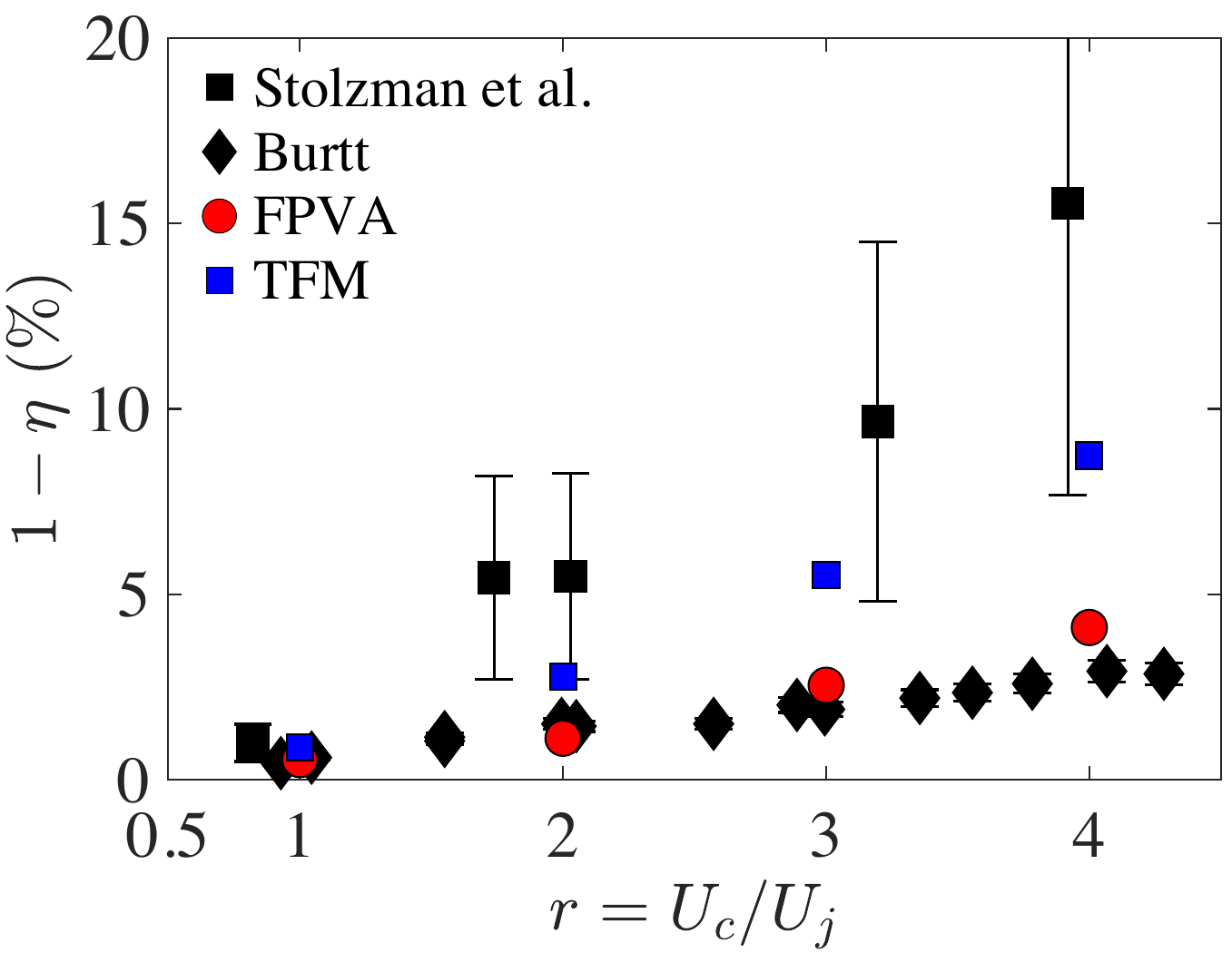}
    \caption{Comparison of combustion inefficiency, $1-\eta$, predicted
    using FPVA and TFM with experimental measurements from the wind-tunnel
    study of Burtt \cite{burtt2014efficiency} and outdoor flare measurements of
    Stolzman et al. \cite{stolzman2025experimental,stolzman2026effects}. Error bars indicate the reported
    experimental uncertainty.}
    \label{fig:jicf_ce_validation}
\end{figure}

The predicted combustion efficiencies are compared with available
experimental measurements in Fig.~\ref{fig:jicf_ce_validation}. Both FPVA
and TFM reproduce the experimentally observed increase in combustion
inefficiency with increasing crosswind-to-jet velocity ratio. This nonlinear deterioration in flare performance with
increasing crosswind has previously been observed in both controlled
wind-tunnel measurements and calorimeter flare experiments
\cite{burtt2014efficiency,stolzman2025experimental}. 

The experimental data exhibit substantial scatter and higher uncertainties at larger velocity ratios, reflecting differences in facility conditions, atmospheric turbulence, and wind variability. The experiments also use continuous pilots, whereas the simulations employ a one-time ignition source. These differences affect point-by-point agreement and indicate that the measurements represent a range of physically observed flare behavior. Regardless of these issues, the growing separation between FPVA and TFM at higher velocity ratios suggests that combustion-model differences become increasingly important as the crosswind strengthens relative to the jet velocity.

\subsection{Premixed branch contributions}
\label{sec:prembranch}
\begin{figure}[!t]
    \centering
    \includegraphics[scale=0.24]
    {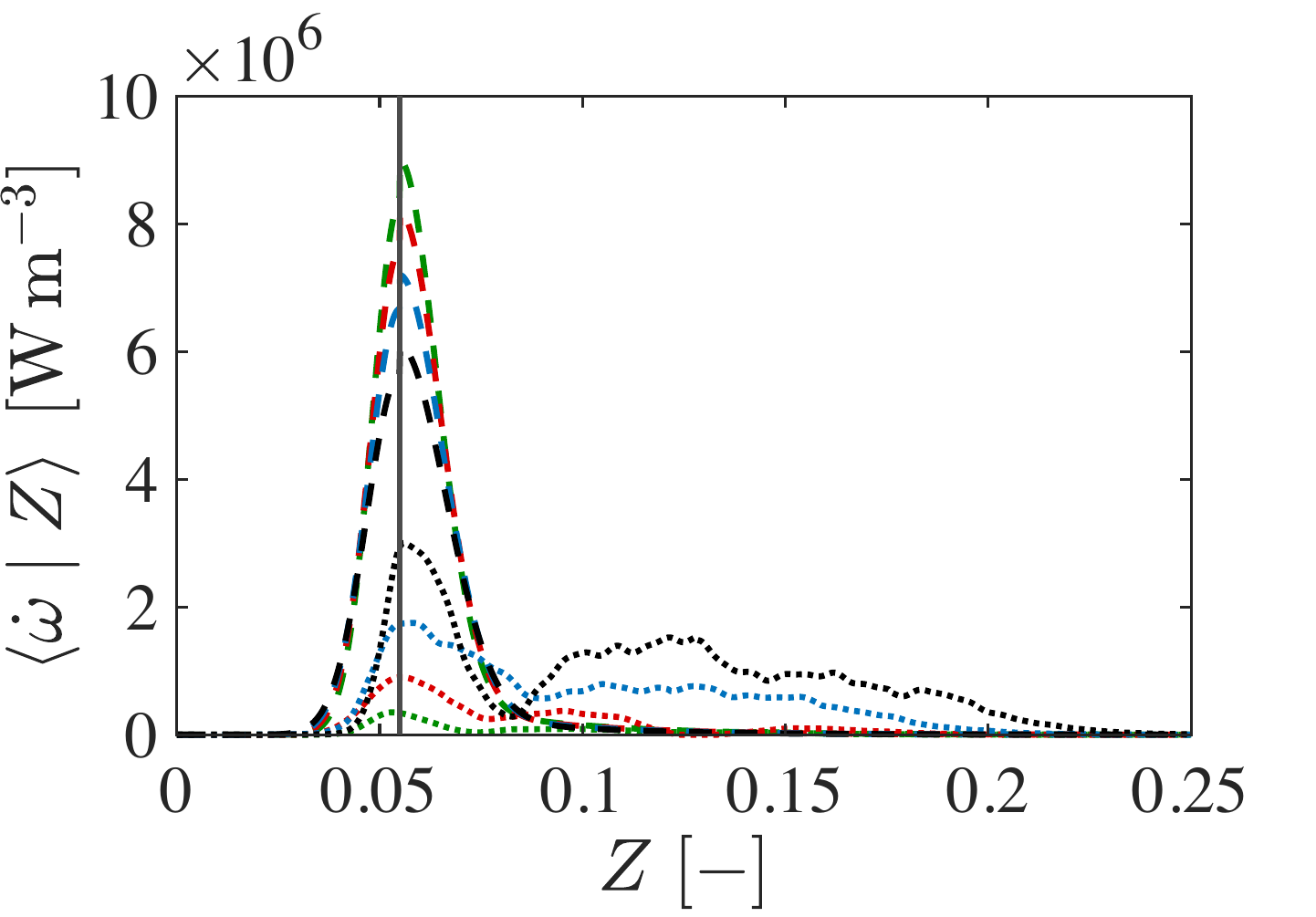}
    \caption {Conditional average of the heat-release rate on mixture fraction, separated into premixed (dotted lines) and diffusion-controlled (dashed lines) branches. Colors denote the crosswind-to-jet velocity ratio: $r=1$ (green), $2$ (red), $3$ (blue), and $4$ (black). $Z$ corresponding to the stoichiometric mixture fraction is highlighted using the vertical line.}
    \label{fig:premixed_branch}
\end{figure}

The contribution of the local combustion regime is further examined by
conditioning the heat-release rate on mixture fraction and separating the
reaction into premixed and diffusion branches using the Takeno Index.
Figure~\ref{fig:premixed_branch} shows that, at the lowest velocity ratio,
$r=1$, heat release is dominated almost entirely by the diffusion
branch, with a strong peak near the stoichiometric mixture fraction,
$Z_{st}$. As the velocity ratio increases, a distinct premixed contribution progressively develops and increases in magnitude. This behavior becomes particularly evident at $r=3$ and $4$, where the premixed branch spans a broader range of mixture fractions and contributes appreciably to the total heat release. The diffusion-controlled branch remains the primary heat-release source near $Z_{st}$, but increasing crosswind promotes fuel--air mixing before reaction, thereby increasing the fraction of the flame that burns under locally premixed conditions. This trend is consistent
with the growing separation between FPVA and TFM observed at the larger
velocity ratios since the FPVA manifold does not
explicitly capture this premixed reaction branch.

\begin{figure}[!t]
    \centering
    \includegraphics[scale=0.32]
    {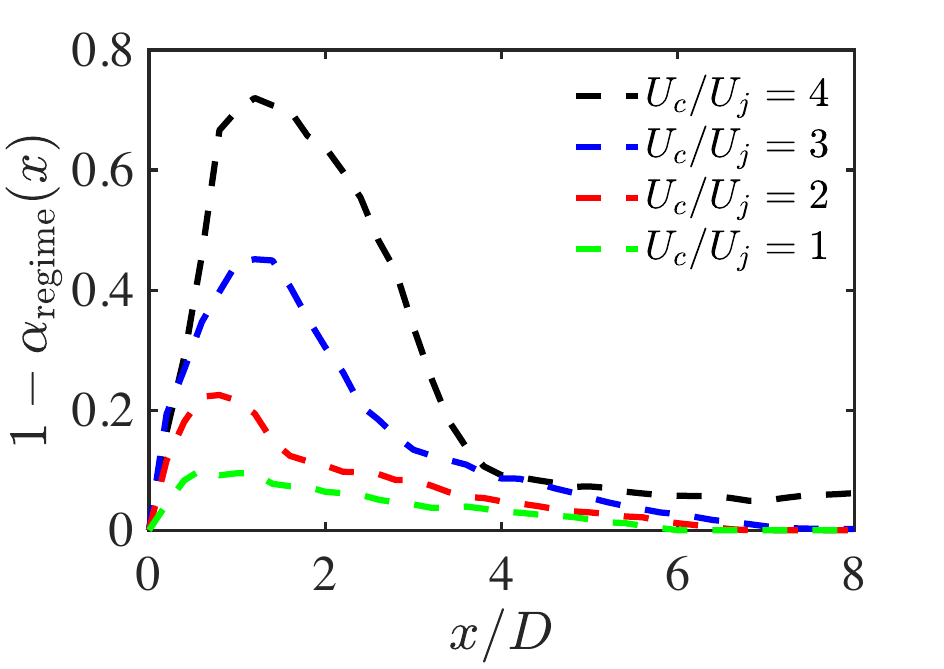}
    \caption{Streamwise variation of the heat-release-weighted premixed
    regime fraction, $1-\alpha_{\mathrm{regime}}$, for
    $U_c/U_j=1$--4. .}
    \label{fig:alpha_regime}
\end{figure}

The relative contributions of the diffusion and premixed branches are
quantified using the regime parameter, $\alpha_{\mathrm{regime}}$. The
mean heat release is separated using the Takeno Index, where
$\dot{\Omega}_{D}=\int_V \dot{\omega}|_{\mathrm{TI}=-1}\,{\rm d}V$ and
$\dot{\Omega}_{P}=\int_V \dot{\omega}|_{\mathrm{TI}=1}\,{\rm d}V$ denote the
diffusion and premixed heat-release contributions, respectively. The
regime parameter is then defined as
$\alpha_{\mathrm{regime}}=
\overline{\dot{\Omega}}_{D}/
(\overline{\dot{\Omega}}_{P}+\overline{\dot{\Omega}}_{D})$. Here, $\alpha_{\mathrm{regime}}$ represents the fraction of total 
heat release associated with the diffusion branch. Consequently,
$1-\alpha_{\mathrm{regime}}$ provides the complementary contribution
from the locally premixed branch,

Figure~\ref{fig:alpha_regime} shows the spatial distribution of $1-\alpha_{\mathrm{regime}}(x)$, which represents the fraction of 
local heat release associated with premixed combustion as a function of the axial distance. The premixed
contribution is concentrated primarily in the near-field and increases
strongly with velocity ratio. At $r=1$,
$1-\alpha_{\mathrm{regime}}$ remains relatively small, indicating that
the flame is predominantly diffusion controlled. As $r$ increases,
the premixed contribution becomes progressively larger, reaching a
maximum near $x/D\approx1-2$. At $r=4$, a substantial fraction
of the near-field heat release is therefore associated with the premixed
branch.

Farther downstream, $1-\alpha_{\mathrm{regime}}$ decreases for all
velocity ratios, indicating a transition toward increasingly
diffusion-dominated combustion. These results are consistent with the
conditioned heat-release profiles shown in Fig~\ref{fig:premixed_branch}, which show that increasing crosswind
promotes the development of a stronger premixed branch near the nozzle,
while diffusion combustion remains dominant over the downstream portion
of the flame.

\begin{table}[!t]
\caption{Summary of combustion efficiency, mean premixed heat-release contribution, and flame length for FPVA and TFM at different velocity ratios.}
\label{tab:premixed_branch_summary}
\centering
\footnotesize
\setlength{\tabcolsep}{3.5pt}
\renewcommand{\arraystretch}{1.15}
\resizebox{\columnwidth}{!}{%
\begin{tabular}{c|cc|cc|cc}
\hline
\multirow{2}{*}{$U_c/U_j$}
& \multicolumn{2}{c|}{$\eta~(\%)$}
& \multicolumn{2}{c|}{$1-{\alpha}_{\mathrm{regime}}$}
& \multicolumn{2}{c}{Flame length (m)} \\
& FPVA & TFM & FPVA & TFM & FPVA & TFM \\
\hline
1 & 99.41 & 99.13 & 1.00 & 0.99 & 1.61 & 1.42 \\
2 & 98.82  & 97.21 & 1.00 & 0.98 & 1.46 & 1.26 \\
3 & 96.54 & 94.47 & 1.00 & 0.92 & 1.29 & 1.07 \\
4 & 94.12 & 91.29 & 1.00 & 0.81 & 1.11 & 0.91 \\
\hline
\end{tabular}%
}
\end{table}


Table~\ref{tab:premixed_branch_summary} summarizes the combustion efficiency, regime contribution, and flame length predicted by FPVA and TFM. At $r=1$, the combustion efficiencies differ by only $0.28$ percentage points, but this gap increases to $2.83$ percent at $r=4$. Across the same range, $\eta$ decreases from $99.41\%$ to $94.12\%$ for FPVA and from $99.13\%$ to $91.29\%$ for TFM. The divergence is accompanied by a stronger change in the combustion regime predicted by TFM, with $1-\alpha_{\mathrm{regime}}$ increasing from $0.01$ to $0.19$ as $r$ increases from 1 to 4.

The flame length, defined using the time-averaged $T=800~\mathrm{K}$ isocontour \cite{bradley2016jet}, decreases with increasing crosswind for both models. FPVA predicts a reduction from $1.61$ to $1.11~\mathrm{m}$, while TFM decreases from $1.42$ to $0.91~\mathrm{m}$. Overall, the two closures agree closely at low velocity ratio but diverge progressively as the crosswind increases, with TFM predicting lower combustion efficiency and a shorter flame.

\subsection{Thermochemical-state and strain response}
\label{sec:state_strain_comparison}

\begin{figure*}[!t]
    \centering

    \begin{subfigure}[t]{0.32\textwidth}
        \centering
        \includegraphics[width=\textwidth]
        {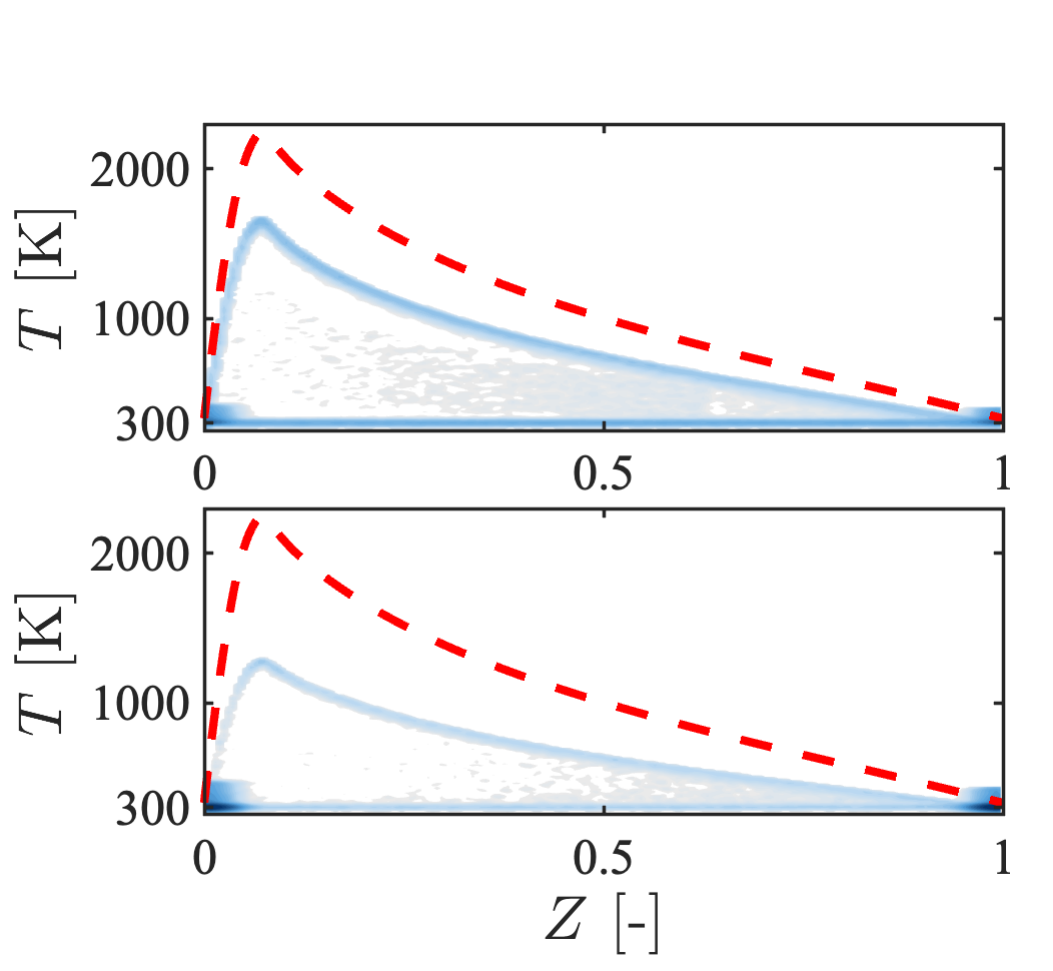}
        \caption{}
        \label{fig:ZT_x1}
    \end{subfigure}
    \hfill
    \begin{subfigure}[t]{0.26\textwidth}
        \centering
        \includegraphics[width=\textwidth]
        {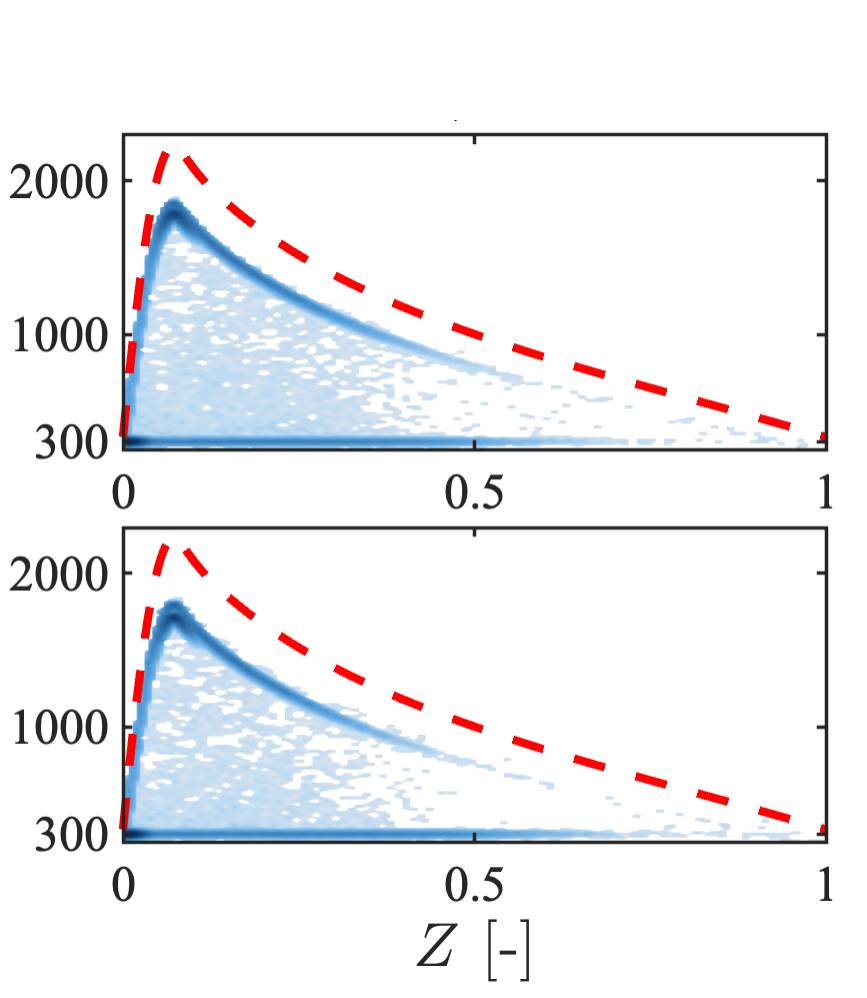}
        \caption{}
        \label{fig:ZT_x6}
    \end{subfigure}
    \hfill
    \begin{subfigure}[t]{0.33\textwidth}
        \centering
        \includegraphics[width=\textwidth]
        {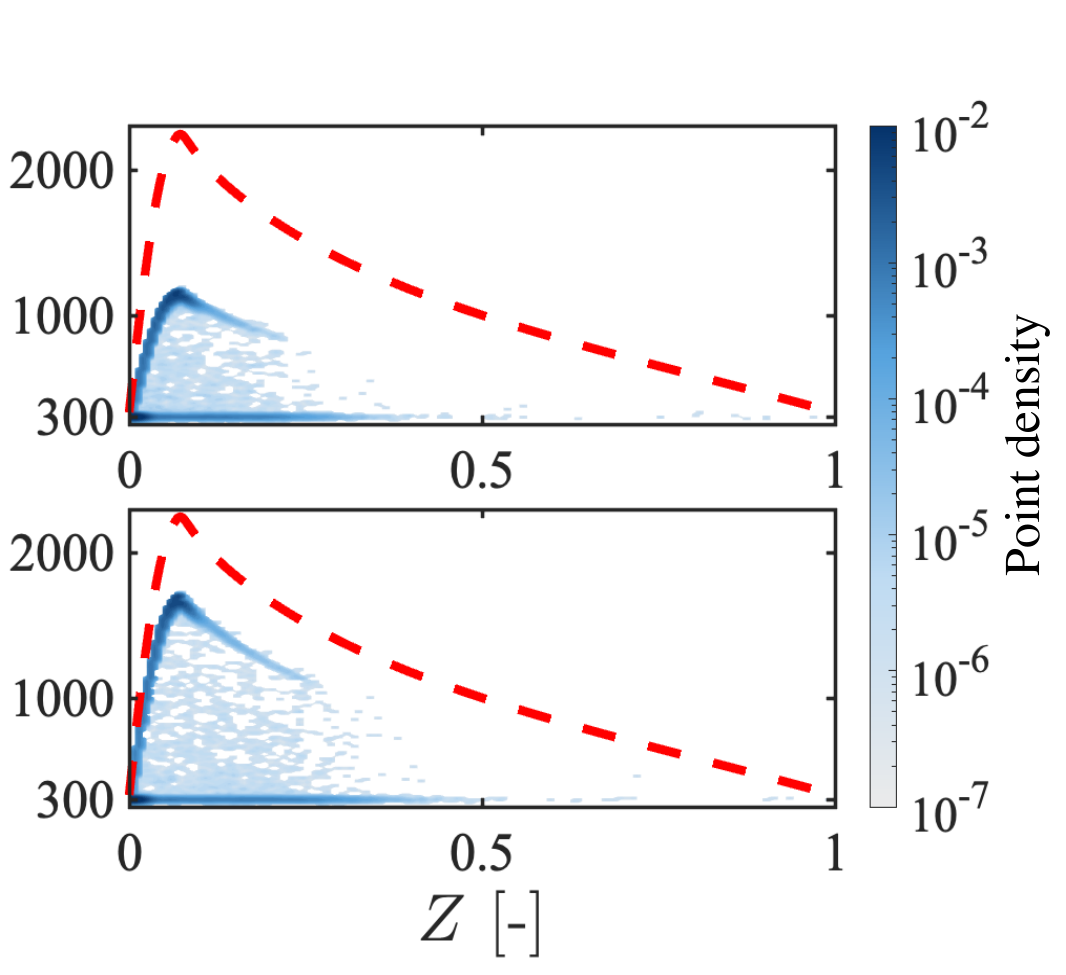}
        \caption{}
        \label{fig:ZT_x12}
    \end{subfigure}

    \caption{Temperature--mixture-fraction distributions predicted by
TFM (top panels) and FPVA (bottom panels) for $r=4$ within three
streamwise sampling volumes centered at $x/D=1$, 6, and 12 corresponding to a) \(0.5\le x/D\le1.5\), b) \(5.5\le x/D\le6.5\), and c) \(11.5\le x/D\le12.5\), respectively. Each
distribution includes all computational cells within the indicated
streamwise interval and across the full cross-stream extent of the domain.
The dashed red line denotes the corresponding FlameMaster
counter diffusion-flame reference solution for a fully burned mixture at equilibrium.}
    \label{fig:ZT_comparison}
\end{figure*}

To further examine the differences between FPVA and TFM, the predicted thermochemical states are compared in temperature--mixture-fraction space. Figure~\ref{fig:ZT_comparison} presents the distribution of LES states within three streamwise sampling volumes centered at \(x/D=1\), 6, and 12, corresponding to \(0.5\le x/D\le1.5\), \(5.5\le x/D\le6.5\), and \(11.5\le x/D\le12.5\), respectively. Each sampling volume includes all computational cells within the corresponding streamwise interval and across the full cross-stream extent of the domain. The upper panels correspond to TFM and the lower panels to FPVA. The dashed red line denotes the one-dimensional solution obtained using FlameMaster. This solution represents the fully burnt flamelet solution at equilibrium and provides a reference \(T(Z)\) relation against which the LES thermochemical states are compared.

The largest differences between the two models are observed in the
near-field. Within $0.5\leq x/D\leq1.5$, TFM populates a broader range
of elevated-temperature states than FPVA, consistent with the stronger
near-nozzle reaction and premixed contribution identified previously.
At $5.5\leq x/D\leq6.5$, the two distributions coincide as the flame becomes increasingly diffusion dominated. Farther
downstream, within $11.5\leq x/D\leq12.5$, the distributions diverge again, with FPVA retaining a broader population of high-temperature states than TFM. Thus, although the two approaches predict similar thermochemical behavior in portions of the diffusion-dominated flame, near-field differences persist and influence the downstream plume.

\begin{figure*}[!t]
    \centering
    \includegraphics[width=0.95\textwidth]
    {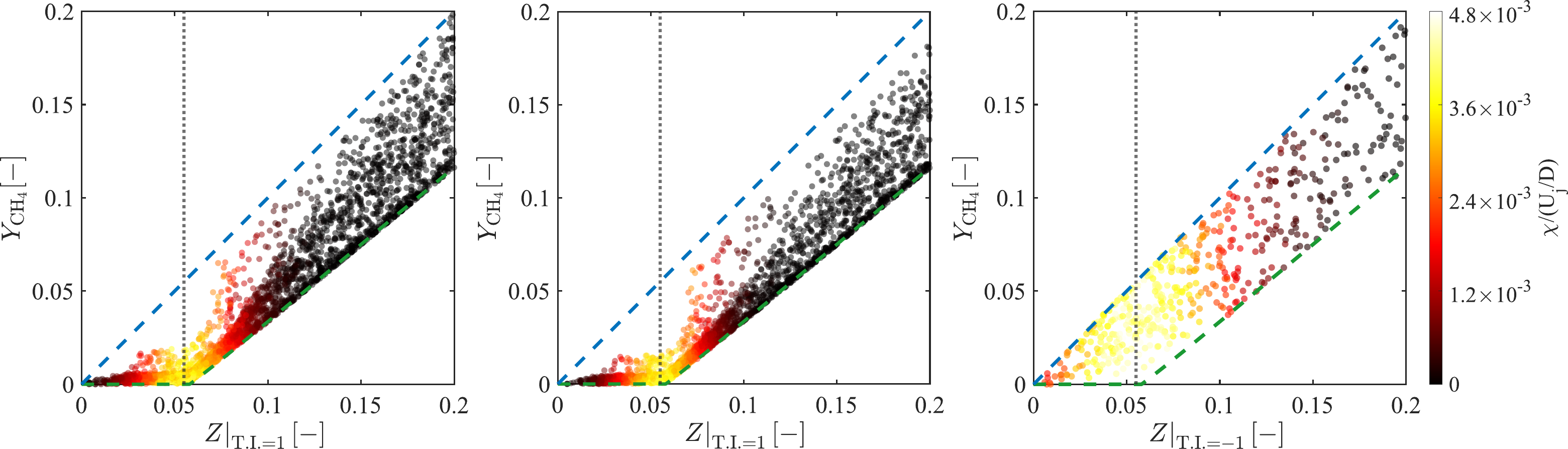}
    \caption{Methane mass fraction as a function of mixture fraction for
    the different combustion regimes in the JICF with $r=4$. From left to right, the panels show
    the TFM diffusion branch ($\rm{TI}=-1$), FPVA diffusion branch ($\rm{TI}=-1$),
    and TFM premixed branch ($\rm{TI}=1$). LES states are colored by the
    normalized scalar dissipation rate, $\chi/(U_j/D)$. The FlameMaster reference for a fully burned mixture at equilibrium (green line), the unburned-mixture relation $Y_{\mathrm{CH_4}}=Z$ (blue line), and the stoichiometric mixture fraction (vertical black line) are shown for comparison.}
    \label{fig:strain_regime}
\end{figure*}

The local strain associated with the two combustion regimes
is shown in Fig.~\ref{fig:strain_regime}. The first two panels represent the
diffusion-classified states predicted by TFM and FPVA, respectively,
while the third panel represents the premixed branch predicted by TFM.
The diffusion-classified states predicted by FPVA and TFM are broadly
similar, whereas the TFM premixed branch exhibits substantially higher
normalized scalar dissipation, defined as $\chi = 2D_Z |\nabla Z|^2$, which quantifies the rate at which mixture-fraction fluctuations are diffused and is commonly used as an indicator of the local
mixing and strain experienced by the flame. Higher $\chi$ therefore
corresponds to stronger scalar gradients and a greater tendency toward
strain-induced weakening or local extinction. The larger values observed
for the TFM premixed branch indicate that this region experiences a
stronger local strain environment than the diffusion-dominated portion of
the flame as it resides in the strong wake in the leeward side of the nozzle, where the dominant flow recirculation zone is present. The coexistence of elevated scalar dissipation and greater downstream methane concentrations provides a possible explanation for the lower combustion efficiency predicted by TFM. Methane entering these highly strained regions may pass through the reaction zones without sufficient time for complete oxidation. This behavior is consistent with previously identified links between flame instability, elevated strain, and methane leakage~\cite{baranger2024stabilization,gaipl2025combustion,Mohit2026}. 

Together, these findings suggest that the growing discrepancy between FPVA and TFM at high crosswinds ($r>4$) arises primarily from differences in their representation of highly strained premixed regions. TFM directly captures locally premixed combustion and finite-rate effects, including flame stretch and local extinction, resulting in a more prominent premixed branch, greater methane slip, and lower combustion efficiency. 


\section{Conclusions}
\label{sec:conclusions}
This study compared FPVA and a thickened-flame model within a common LES framework across two configurations: the canonical Sandia D flame and a methane jet-in-crossflow representative of a non-assist flare. For the Sandia-D flame, FPVA and TFM produced similar mean predictions,
with larger differences primarily in the RMS quantities.
FPVA achieved grid independence at substantially lower cost, requiring approximately
1000 CPU-hours per flow-through time compared with 2600 CPU-hours for TFM.

For the jet-in-crossflow configuration, TFM
required approximately 30 million elements compared with 15 million for
FPVA and predicted progressively lower combustion efficiency as the
 velocity ratio increased. This coincided with
the emergence of a locally premixed reaction branch captured by TFM but
absent from the present non-premixed FPVA approach.

Regime analysis showed that the diffusion branches predicts by
FPVA and TFM remain broadly similar, whereas the TFM premixed branch
experiences substantially higher scalar dissipation. TFM also predicted stronger near-nozzle recirculation and greater downstream methane,
suggesting a possible link between strained premixed combustion and reduced
efficiency.

Overall, FPVA provides a computationally efficient approach for predicting combustion efficiency trends when the local flame structure remains predominantly non-premixed and consistent with flamelet assumptions. At higher crosswind velocities, increased strain and localized premixing introduce finite-rate effects that may challenge the underlying non-premixed flamelet representation. TFM offers greater flexibility in capturing these combustion-regime transitions and associated methane leakage, albeit at higher computational cost. Thus, model selection should be guided by the local combustion regime and the relative importance of chemical and transport timescales.

\section*{CrediT authorship contribution statement}
\textbf{A. Mohit}: Writing – original draft, Visualization, Validation, Software, Methodology, Investigation, Formal analysis, Data curation. 
\textbf{M. Wooldridge}: Writing – review and editing, Supervision, Project administration, Funding acquisition, Formal analysis.
\textbf{J. Capecelatro}: Writing – review and editing, Supervision, Project administration, Funding acquisition, Formal analysis.

\section*{Declaration of competing interest}

The authors declare that they have no known competing financial interests or personal relationships that could have appeared to influence the work reported in this paper.

\section*{Acknowledgments}
This work was supported in part by the Advanced Research Projects Agency-Energy (ARPA-E), U.S. Department of Energy, under Award No. DE-AR0001534. Computational resources and services were provided by Advanced Research Computing (ARC) at the University of Michigan and the San Diego Supercomputer Center (SDSC) through NSF ACCESS Award No. MCH260119.

\FloatBarrier

\bibliographystyle{cnf-num}
\bibliography{CNF-template/references_corrected_capitalization}

\end{document}